\documentclass[iop]{emulateapj}
\newcommand{\oii}{[\ion{O}{2}]}
\newcommand{\oiii}{[\ion{O}{3}]}
\newcommand{\hb}{H$\beta$} 
\newcommand{\ha}{H$\alpha$}
\newcommand{\lya}{Ly$\alpha$} 

\newcommand{\nii}{[\ion{N}{2}]}

\begin{document}
\title{The metallicity evolution of low mass galaxies: New constraints at intermediate Redshift\altaffilmark{1}} 
\author{Alaina Henry\altaffilmark{2,3,4},  Crystal L. Martin\altaffilmark{2}, Kristian Finlator\altaffilmark{2,5}  \& Alan Dressler\altaffilmark{6}
}
\altaffiltext{1}{Some of the data presented herein were obtained at the W.M. Keck Observatory, which is operated as a scientific partnership among the California Institute of Technology, the University of California and the National Aeronautics and Space Administration. The Observatory was made possible by the generous financial support of the W.M. Keck Foundation.}
\altaffiltext{2}{Department of Physics, University of California, Santa Barbara, CA 93106}
\altaffiltext{3}{Astrophysics Science Division, Goddard Space Flight Center, Code 665, Greenbelt, MD 20771; alaina.henry@nasa.gov}
\altaffiltext{4}{NASA Postdoctoral Program Fellow} 
\altaffiltext{5}{Hubble Fellow} 
 \altaffiltext{6}{Carnegie Observatories, 813 Santa Barbara Street, Pasadena, CA 91101}

\begin{abstract} 
We present abundance measurements from 26 emission-line selected galaxies  at 
$z\sim 0.6-0.7$.  By reaching stellar masses as low as $10^{8}$ M$_{\sun}$, these observations provide the first
measurement of the intermediate redshift mass-metallicity (MZ) relation below $10^{9}$ M$_{\sun}$.  
For the portion of our sample above $M > 10^{9}$  M$_{\sun}$ (8/26 galaxies), we find good agreement with previous measurements of the intermediate redshift MZ relation.   
Compared to the local relation, we measure an evolution that corresponds to a  0.12 dex decrease in oxygen abundances at intermediate redshifts. This result confirms the trend  that metallicity evolution becomes more significant towards lower stellar masses, in keeping with a downsizing scenario where low mass galaxies evolve onto the local MZ relation at later cosmic times.  We show that these galaxies follow the local fundamental metallicity relation, where objects with higher specific (mass-normalized) star formation rates (SFRs) have lower metallicities.       
Furthermore, we show that the galaxies in our sample lie on an extrapolation of the SFR-M$_{*}$ relation (the star-forming main sequence). 
Leveraging the MZ relation and star-forming main sequence (and combining our data with higher mass measurements from the literature), we test models that assume an equilibrium between mass inflow, outflow and star formation.   
We find that outflows are required to describe the data.
By comparing different outflow prescriptions, we show that momentum driven winds can describe the MZ relation; 
however, this model under-predicts the amount of  star formation in low mass galaxies.    This  disagreement may indicate 
that preventive feedback from gas-heating has been overestimated, or it may signify a more fundamental deviation from the equilibrium assumption. 
 \end{abstract}

\section{Introduction} 
The balance between gaseous inflows, outflows, and star formation is a critical
frontier in our understanding of galaxy evolution. 
Feedback caused by stellar winds, supernovae, and supermassive black holes 
is often used to explain a variety of observations, from luminosity 
and stellar mass functions, to the enrichment and reionization of the intergalactic medium.    
However,  a complete physical picture of these feedback processes  (i.e.\ \citealt{Murray, Hopkins}) is still debated.   
Continued efforts to provide new observational tests are essential.   

The correlation between galaxy stellar masses and gas-phase metallicities (the MZ relation) is one important
probe of star formation feedback.    The galactic outflows that slow star formation, and the inflows that promote it can also alter metallicities.  
On one hand, metal poor  material, when accreted onto a galaxy, can lower the  metallicity.  
On the other hand, supernova driven winds may remove metal enriched material from galaxies.  It has been known for many years that models which ignore inflows and outflows (i.e.\ closed-boxes), fail to reproduce
observed abundance patterns in galaxies \citep{vdB62}.   Indeed, in recent years,  both analytical and numerical 
models have shown that outflows (and sometimes gas accretion) are needed to explain the MZ relation  \citep{Tremonti, Dalcanton07, Brooks07, Finlator08, Erb08, Dave11b, Dave12, PS11, Dayal, Lilly13}.   However, to date, studies  have not converged on the 
properties (rates, kinematics, metal enrichment, halo mass dependence, and redshift evolution) of these gaseous flows. 

\begin{deluxetable*} {ccccccccc}[!ht]
\tablecolumns{9}
\tablecaption{DEIMOS Followup Observation Summary} 
\tablehead{
\colhead{Date}  &  \colhead{Mask Name}  & \colhead{Mask RA}  & \colhead{Mask Dec} &\colhead{Mask PA}  & \colhead{Slit PA} & \colhead{Slit Widths} & \colhead{Exposure Time}  & \colhead{Seeing}   \\ 
  &   &  (J2000)  & (J2000) & (degrees)  & (degrees) & (\arcsec)   & (hours) & (\arcsec)  
} 
\startdata
27 January 2011& D &  10:00:22.97 &    02:09:28.9  & 85  & 90  & 1.5 &  6.5		&	0.6  \\
28 January 2011&  F &    10:01:11.46 &     02:10:27.4 & 106  & 90  & 1.5 &  6.3		& 0.8		 \\
22 January 2012 & M &  10:00:24.25 &    02:05:19.8  &  85  &   90 &  1.5 & 4.9    &  1.0    \\
22/24 January 2012 &  L &  10:00:22.56     &  02:15:29.8 &  13 & 0  &    1.2 & 5.3   & 0.9  \\ 
23/24 January 2012 & Q  & 10:00:28.96   &  02:20:05.1 & 95  & 90 &  1.5 &   6.8 &  1.0   
\enddata
\label{obstable} 
\tablecomments{Coordinates,  position angles (PAs), exposure times, and  seeing are given for each observed mask.  On masks D, F, M, and Q, the 
slit PAs and slit-widths were matched to the  IMACS venetian blind spectroscopic  data (described in \citealt{Dressler11} and \citealt{Henry12}).  For mask L, the slits were rotated 90 degrees relative to the 
search data.   }
\end{deluxetable*}

One avenue to better understand the physics that governs the MZ relation is to study low mass galaxies (e.g.\ \citealt{Lee, Zhao, Wuyts, Berg}). 
In these systems, galactic winds are especially effective at escaping the 
gravitational potential of their hosts, regulating star formation and  
enabling enrichment of the intergalactic medium \citep{Opp09,Kirby11}.     Hence, observational constraints
on low mass galaxies offer some of the most stringent tests of galaxy formation models.  
 
Outside the local universe, the MZ relation is poorly constrained at stellar masses below $10^{9}$ M$_{\sun}$ (and $10^{10}$  $M_{\sun} $ for $z>1$).   
While large spectroscopic surveys  \citep{vvds, Lilly, Newman}   have enabled abundance measurements  of statistical samples out to $z\sim1$,  these surveys are typically limited  to $R \la 24$.  Hence, these intermediate-redshift MZ relations have been  derived for $M > 10^{9}$ M$_{\sun}$ \citep{Lilly03, Savaglio, Lamareille, Zahid11, Cresci, Moustakas11}.   Nevertheless, in the case of the Cosmic Origins Survey (COSMOS),  the extensive  broad and intermediate-band photometry allows reliable mass constraints an order of magnitude lower at intermediate redshifts.   Therefore, spectroscopic followup of fainter galaxies can significantly extend the  intermediate redshift MZ relation.

In this paper we use an emission line selected sample to place new constraints on the low mass 
end of the MZ relation at $z\sim 0.6-0.7$.   By drawing our sample primarily from 
the ultra-faint emission line objects that we have previously identified with blind spectroscopy in the COSMOS field \citep{Martin08, Dressler11, Henry12}, we obtain oxygen abundances for galaxies with stellar masses of $10^{8} ~ {\rm M}_{\sun} \la {\rm M} \la 10^{10}~ {\rm M}_{\sun}$.    
In this manner, we provide the first constraints on the low mass evolution of the MZ relation, reaching stellar masses
that are comparable to the limiting mass of local SDSS samples.

This paper is organized as follows: in \S \ref{obs} we describe our spectroscopic observations and the COSMOS imaging data
that we use.  In \S \ref{analysis}  we describe our emission line measurements and  stellar mass derivations.  Then, in \S \ref{calc_metal} we calculate the oxygen abundances of galaxies in our sample, discussing the various diagnostics that have been proposed to break the degeneracy of the double-valued R23 metallicity indicator \citep{Pagel}.  
In \S 5 and \S 6 we compare our MZ relation to previous derivations, and investigate the presence of a mass-metallicity-SFR fundamental plane.  Finally, we 
compare to theoretical predictions of the MZ relation in \S 7.   In this paper we use AB magnitudes, a \cite{Chabrier} initial mass function,  
and a $\Lambda$CDM cosmology with $\Omega_M$ = 0.3, $\Omega_\Lambda = 0.7$ and $H_0$ = 70 
km s$^{-1}$ Mpc$^{-1}$.    Throughout the text we report measurements of doublet lines: \oii\ $\lambda \lambda 3727, 3729$ and \oiii\ $\lambda \lambda 4959, 5007$.  
For the sake of brevity, we use the notation ``\oii'' and ``\oiii'' to refer to both lines in the doublet, or, when 
appropriate, the sum of their fluxes

\section{Observations} 
\label{obs} 

\subsection{Target Selection \& Followup Spectroscopy} 
The emission line galaxies in the present sample were initially identified as part of our multislit narrowband spectroscopic survey.  
The observations are presented in detail in \cite{Martin08}, \cite{Dressler11}, and \cite{Henry12}.   In brief, our design uses the Inamori-Magellan Areal Camera and Spectrograph (IMACS; \citealt{imacs})  on the 6.5m Magellan Baade Telescope at Las Campans Observatory.  We used a venetian blind slit mask,  and narrowband filter centered in the 8200\AA\ OH airglow free window.  This method allows for the efficient selection of emission line 
galaxies, and reaches fluxes as low as 2.5$\times 10^{-18}$ erg s$^{-1}$ cm$^{-2}$-- a factor of five fainter than narrowband
imaging surveys (e.\ g.\  \citealt{Kashikawa11}).     
 
\begin{figure*} 
\plotone{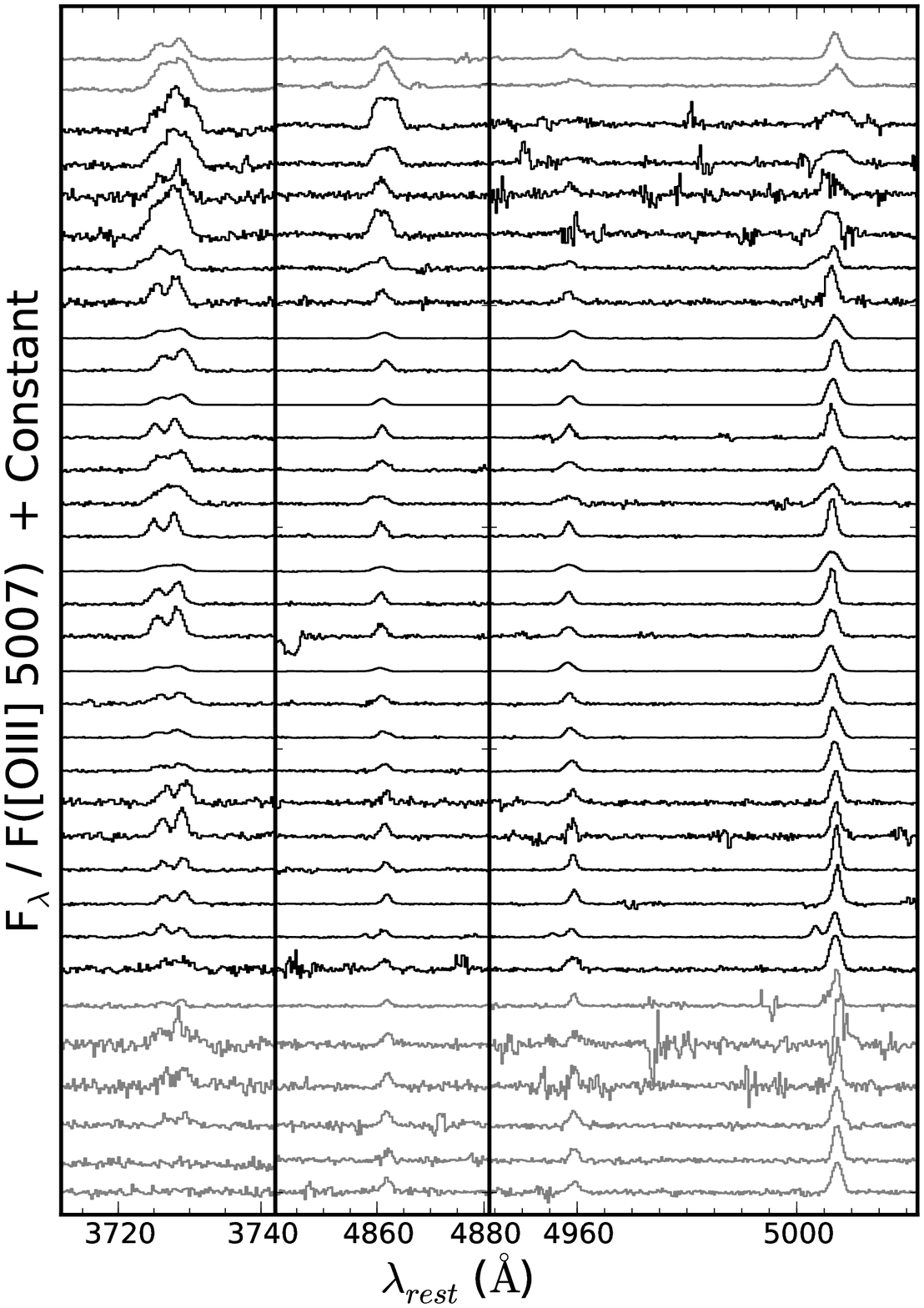} 
\caption{ The spectra of 34 galaxies with complete coverage of  the R23 metallicity indicator (\oiii + \oii  / \hb; \citealt{Pagel})  are shown.    The spectra of 26 objects that comprise our MZ sample are shown in black, and are sorted by stellar mass with higher masses at the top (the same as their appearance in Tables \ref{meas} and \ref{derived}).  The two AGN candidates are shown in grey at the top, and also continue the mass-ordered scheme.  The Class 3 objects with ambiguous optical counterparts are shown at the bottom (in grey) in the same order as their appearance in the tables.    All spectra are normalized by their flux in [OIII] $\lambda 5007$,  so it is  apparent that higher mass galaxies have relatively stronger \hb\ and \oii\ emission.   The strong ``absorption'' trough near 4845 \AA\ in one object (980802) is a detector artifact. } 
\label{allspec} 
\end{figure*}

Because of the faintness of the emission lines, and the narrow bandpass in our search data, we use followup spectroscopy with DEIMOS \citep{Faber}
to identify the redshifts of the IMACS-detected galaxies.  The primary goal of these observations was to confirm \lya\ emitting galaxies and measure the faint-end slope of the \lya\ luminosity function at $z=5.7$ \citep{Henry12}.    
However, for the foreground galaxies at  $z\sim 0.6-0.7$, these followup observations contain the \oiii $\lambda \lambda 4959, 5007$, \oii $\lambda \lambda 3727, 3729$, and \hb\ emission lines that comprise
the R23 metallicity indicator \citep{Pagel}.  Three classes of these objects were included on followup slit-masks:
\begin{enumerate} 
 \item Emission line galaxies for which continuum in the search data ruled out the  \lya\ identification were chosen as MZ targets.   To obtain the best possible spectrum, we searched 
the COSMOS photometric redshift catalog \citep{Ilbert09} for nearby galaxies  ($<2$\arcsec) with $z\sim0.6-0.7$, and shifted the slit-positions 
to coincide with these matches\footnote{Because the COMSOS photometric redshifts include narrowband photometry and use templates that 
include emission lines, the photometric redshifts are often very close to the spectroscopic redshifts for the emission-line selected galaxies in our survey.}.    In total, the final sample of 26 star forming galaxies (described in \S \ref{analysis}) contains 15 objects that meet these criteria. 
\item Additionally, galaxies were drawn from the \oiii\ + \hb\ narrowband excess catalog which we derived in \cite{Dressler11}.    The present star-forming sample contains 11 objects that were selected in this way.  
\item Finally, we also detect R23 from  ultra-faint emission lines which we originally considered \lya\ candidates, but, upon the followup observations described below, we determined that the discovery line was  \oiii\  $\lambda 4959$,  \oiii\ $\lambda 5007$, or \hb.   Six galaxies with measurable R23 fall under this classification.
\end{enumerate} 
In practice, the  objects described in \#3 are difficult to evaluate; their imaged counterparts  are sometimes ambiguous due to their faintness and uncertain position within the blind-search slit.   
Therefore, in this paper we focus on the former two classes. For completeness,  in Tables 2 and 3 we list the subset of these objects where
R23 could be measured; however, we do not consider them further in this paper. 

 The DEIMOS observations of these followup slit-masks were carried out in January 2011 and January 2012.     A summary of
observations for each followup mask is given in Table \ref{obstable}.  In four of the five masks, we used slit-widths and PAs that were matched to the search 
data (1.5\arcsec\ wide and 90\degr\ east of north).  On mask L, in order to better locate objects detected through blind spectroscopy, we used a slit orientation which is orthogonal to the venetian blind search slits.  This method also allowed for narrower slits (1.2\arcsec).  All observations were carried  out using the the GG495 blocking filter with  830G grating.  Under this configuration, we achieved a spectral resolution of 3.7 \AA\  (2.9 \AA) for a source that fills the 1.5\arcsec\  (1.2\arcsec) slit.  The grating angle was chosen to give a central wavelength of 7270 \AA, with blue coverage down to 5500\AA\ for most slits.

 The DEIMOS spectra were reduced using the DEEP2 DEIMOS data reduction pipeline \citep{Cooper}, with an updated optical model 
for the 830G grating (P. Capak, private communication).   The data were flux calibrated using observations of several spectrophotometric 
standard stars, taken through a 1.5\arcsec\ slit at the parallactic angle. The stars used were G191B2B,  GD50, Feige 66, Feige 67, and Hz 44 \citep{MG90, Oke}, and the data for each was taken from the ESO spectrophotometric standard star database.   Sensitivity functions derived from these stars differ systematically between observations, with offsets up to 30\%.  
 After shifting the lower sensitivity functions to match the highest one, the observations agree at the 2-3\% level, indicating an excellent 
relative calibration. 

Finally, we have verified that the effects of differential atmospheric refraction have a negligible impact on our  measured line flux ratios.    
In order to quantify possible differential slit losses, we calculate the component of the atmospheric refraction that falls perpendicular
to the slit in each observation (frame) of each mask.  The worst case occurs on mask L, where the slits are narrowest and the slit-PA  
was the furthest from parallactic.   For these data, the refraction (perpendicular to the slit) between 6000 \AA\ (near \oii)  and 8000 \AA\ (near \oiii\ and \hb)  ranges from 0.04 to 0\farcs33 \citep{Fillipenko}.    Therefore, since the guiding was done in the R-band (in between the observed wavelengths of the emission lines), 
we estimate the slit-losses on the red and blue parts of 
our spectra differ by no more than 4\% (for 1\arcsec\ seeing).  In most frames, the effect is even smaller, so we conclude that no systematic correction is needed 
to interpret our emission line ratios.

\subsection{COSMOS Imaging Data} 
In order to derive stellar masses (\S \ref{analysis}), we use the wide range of imaging data provided by the COSMOS team.  
  A more detailed description of the catalogs can be found  in \cite{Capak07} and \cite{Ilbert09}. 
In summary, the photometric data include GALEX imaging, a wealth of ground-based, broad-band optical and near-infrared imaging,
and the four {\it Spitzer}/IRAC bands.   Furthermore, our SED fits are improved by the inclusion of intermediate-band imaging, and updated 
Subaru/SuprimeCam $z'$-band data (from observations made with the new, fully depleted, red-efficient CCDs). 
Since the optical and near-infrared photometry are performed in 3\arcsec\ apertures on data with homogenized PSFs  (1.5\arcsec\ FWHM), we apply a point source aperture correction of -0.28 magnitudes to all optical and near-infrared bands.   For the IRAC data we use 3.8\arcsec\ diameter apertures, and apply point 
 source aperture corrections of -0.29, -0.33, -0.51, and -0.59 magnitudes to bands one through four.   No aperture corrections are applied to the 
 GALEX data, as these magnitudes were derived from fits to the PSF.  Finally,  we note that the recommended zero point offsets have been applied \citep{Capak07, Ilbert09}, and a Galactic foreground extinction correction is made using the  \cite{Cardelli} extinction curve with E(B-V) = 0.019.

\begin{figure*}[!ht]
\plotone{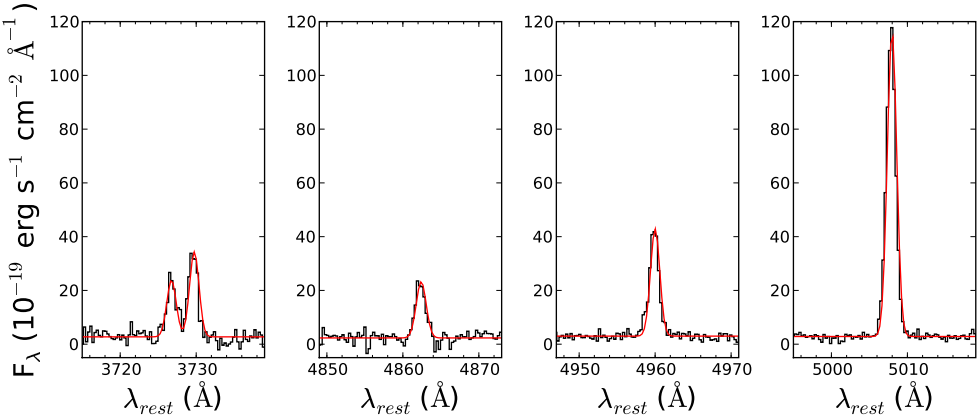} 
\caption{The DEIMOS spectrum of object 54.5+4-0.18 measures the R23 oxygen abundance indicator with high signal to noise.  The Gaussian fits
used to measure the emission line fluxes are shown in red.  From left to right, the lines shown are \oii\ $\lambda \lambda 3727, 3729$,  \hb, \oiii\ $\lambda 4959$,  and \oiii\ $\lambda 5007$.   
This spectrum is representative of the fainter, lower mass galaxies in our sample (see Tables \ref{meas} and \ref{derived}).} 
\label{example_spec} 
\end{figure*}

\section{Analysis} 
\label{analysis} 
\subsection{Emission Line Measurements} 
In total, we identify 47 galaxies at $z\sim 0.6-0.7$; of these 34
have complete coverage of the R23 (\oiii + \oii  / \hb; \citealt{Pagel}) metallicity indicator.   Their 
spectra are shown in Figure \ref{allspec}.  The remaining 13 galaxies at these redshifts either have essential lines lost in the OH  airglow spectrum, or are
 class \#3 objects (described above), where the observed spectrum is especially faint because the galaxy may not be centered in the slit.     
 Only six class \#3 objects (without mass measurements) have R23 measurements; their spectra are shown in grey at the bottom of Figure \ref{allspec}. 
 Excluding these objects leaves 28 galaxies with  both mass and R23 constraints.   In \S 3.3 we show that two of these galaxies may contain active nuclei, so we adopt a final sample that includes the remaining 26 star-forming objects. 

Emission line fluxes are measured for all of the $z\sim 0.6-0.7$ galaxies that we identified. 
Gaussian profiles are fit to each line, and uncertainties are determined through a Monte Carlo simulation where noise 
is added to the spectrum and the fits are repeated.  For \oii, the two doublet components are fit simultaneously. As an example, Figure \ref{example_spec}  shows a typical spectrum and the  Gaussian fits to the emission lines.

To facilitate a correction for  \hb\ stellar absorption, we also measure emission line equivalent widths. 
Because it can be difficult to assess the uncertainty on the continuum flux density when it is measured spectroscopically, we compare two methods.  
On one hand, we measure the equivalent widths directly from  the spectra.   In 25/28 objects in our  sample, 
we detect continuum, although it is sometimes weak.   As a second method, we compare the emission line fluxes from the DEIMOS data to the 
continuum under the emission line from the SED fit (described below).    After accounting for a systematic offset (because the emission line fluxes are 
subject to slit losses but the SED fits use total fluxes), the equivalent widths measured by these different techniques agree to within 50\%.   We adopt this level of uncertainty on the \hb\  equivalent widths.

Finally, having measured the fluxes and equivalent widths of our emission lines, we apply a correction for the stellar absorption.  Because the stellar absorption component is much broader than the  emission line, this correction depends on the spectral resolution.  
At the resolution of our data, the correction is approximately 1 \AA\  in equivalent width \citep{CB08, Zahid11}.  While the amount of  absorption also depends slightly on the age of the stellar population, the main source of uncertainty in this correction is from the observed \hb\  emission equivalent width, as discussed above.    We propagate this 
error in our \hb\ fluxes  and the metallicities that we infer in \S \ref{calc_metal}.

\subsection{Stellar Masses, SFRs, and dust constraints from SED fitting}
Stellar masses are determined by fitting template stellar populations to the broad- and intermediate-band photometry described in \S  2. 
 For this task, we use FAST (Fitting and Assessment of Synthetic Templates; \citealt{FAST}).    The choice of population synthesis templates can have an  important impact on the properties derived from SEDs, because the contribution from the (infrared-bright) thermally pulsing asymptotic giant branch (TP-AGB) stars differs from model to model.  The most widely used templates are Bruzual \& Charlot (2003; BC03),  \cite{M05}, and Charlot \& Bruzual (2007; \citealt{Bruzual07});  the latter two options provide a larger contribution from TP-AGB stars (and correspondingly smaller masses).   Nevertheless, the proper contribution from TP-AGB stars remains a subject of debate  \citep{Kriek10, Conroy10, Zibetti}. 
 Therefore, in order to facilitate comparisons with the literature, we derive stellar masses using the BC03 models.  We have verified that the Charlot \& Bruzual and Maraston et al. models produce
stellar masses that are systematically smaller by approximately 0.1 dex.

In order to assure accurate stellar population constraints, it is also important to account for emission line contribution to our SEDs.  
(The stellar synthesis templates that we fit to the SEDs do not contain nebular features.  Failing to remove emission line
contamination can result in incorrectly inferred ages and stellar masses; \citealt{SdB, Atek11}.)   We use our line fluxes to calculate the small contribution from \oii\ to the $r$-band, and \oiii\ and \hb\ to the $i$-band.    For the galaxies in our sample, we find that 1-5\% of the $r-$ band
light and 2-15\% of the $i-$band light can be attributed to emission lines.    These contributions are subtracted from the $r$ and $i-$ band continuum flux densities.  
Emission from \ha, on the other hand, falls between the  $z'$ and $J$ bands, so it does not contaminate any of the photometry in the present analysis.    Finally, we consider emission line contamination to the intermediate band photometry.  In these cases, the corrections  will be larger and more uncertain.  Therefore, we exclude the band at
624 nm that includes \oii, and the band at  827 nm that  covers \oiii\ and \hb.  The remaining ten intermediate bands should be relatively unaffected.

The  grid of stellar population parameters that we fit with FAST  include: a set of exponentially declining star formation histories with e-folding times, $\tau$,  ranging from 40 Myr to 10 Gyr;
characteristic stellar population ages ranging from 50 Myr to the age of the universe at $z\sim 0.6-0.7$; and $A_V = 0- 3$ for a \cite{Calzetti} extinction curve.   Additionally, we use a \cite{Chabrier}  IMF with 
 metallicities of 0.004, 0.008, and 0.02 (solar).    Supersolar metallicities are excluded,
 because, as we will show in \S \ref{calc_metal}, the galaxies in our sample mostly have subsolar to solar gas-phase metallicities.  
  The derived  stellar masses, SFRs, and visual extinction values are given in Table \ref{derived}.    Because of the intermediate band photometry, the dust extinction constraints
  exclude much of the allowed parameter space.  Therefore we use the SED-derived dust constraints to correct our emission line ratios, including these uncertainties in our error budget.   In making this correction, we also account for the fact that nebular extinction is 2.3 times higher than stellar extinction on average \citep{Calzetti, Cresci}.   While 
  the precise relation between stellar and nebular extinction  remains controversial \citep{CB08, Cresci, Wofford}, we adopt the ``standard''  \cite{Calzetti} relation to allow comparison with other studies.

 In  \S \ref{fmr_sec}  and \ref{model_sec} we draw conclusions based on the SFRs that we have derived from these SED fits. 
 In order to verify that the SFRs are reliable, we have carried out an assessment of the systematic uncertainties by comparing SFRs derived from  \oii,   \hb, and  the SED fits.   The results, outlined in the Appendix, show that the agreement between the different diagnostics is good; systematic offsets are smaller than 0.2 dex.  This level of uncertainty does not affect our conclusions.

 \subsection{Contamination from AGN} 
\begin{figure} 
 \plotone{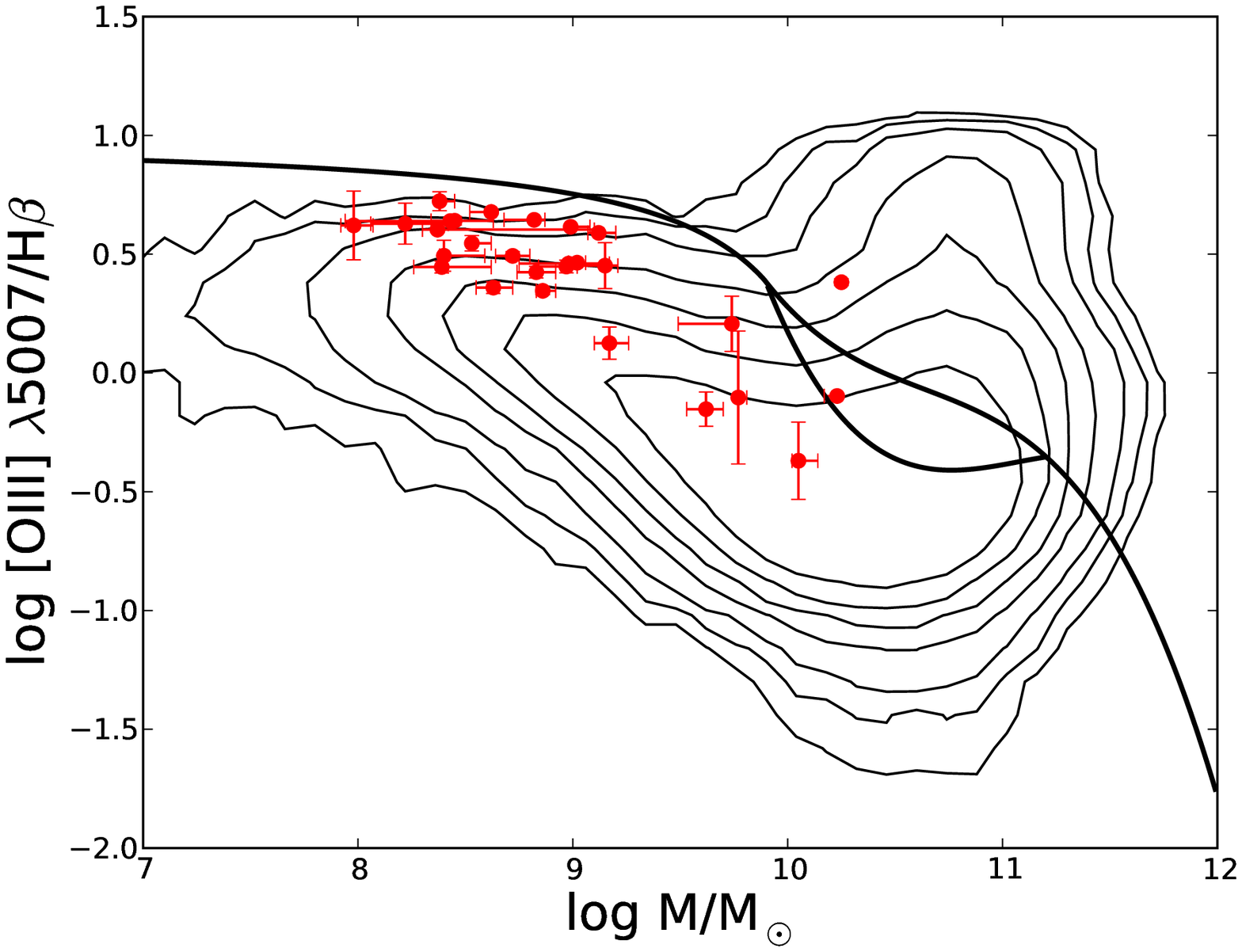} 
 \caption{The \oiii\ / \hb\ ratio as a function of stellar mass can be used to identify AGN, similar to the \oiii/ \hb\ vs. \nii / \ha\ ``BPT'' diagram \citep{BPT, Juneau11}.  Red point show our emission line selected sample, while contours (with levels defined arbitrarily) show the SDSS data. The solid black lines show the demarcation between star-forming galaxies (below and to the left), objects with active nuclei (above and to the right), and composite objects (the region in between the black lines around   $10^{10}$- $10^{11}$ M$_{\sun}$).    The two highest mass objects in our sample, 55.5-3-0.60  and 73.5+5-0.54, fall outside of the $z=0$  star-forming locus, so we identify them as candidate AGN.   } 
 \label{bpt_mex} 
 \end{figure}

 In order to measure the MZ relation of star-forming galaxies, it is important to ensure that the observed emission
 lines do not originate from an active galactic nucleus (AGN).  The typical approach for low redshift galaxies is to use the \oiii/ \hb\ and \nii / \ha\ 
 emission line ratios (the canonical BPT diagram;   \citealt{BPT}).    However, for the redshift of our sample, \ha\ and \nii\ fall at infrared wavelengths.  
 Therefore, in lieu of followup spectroscopy, we turn to an alternate diagnostic:  the Mass-Excitation (MEx) diagram \citep{Juneau11}.  This approach, shown in Figure \ref{bpt_mex}, uses stellar mass as a proxy for  the \ha/\nii\ $\lambda 6583$ ratio (relying on the MZ relation).   The resulting diagram appears qualitatively similar to the traditional BPT diagram, with AGN falling towards the top and right of the plot.  In this figure, the data for our galaxies (red points) are compared to that from the SDSS (contours),  and the solid black line demarcates the difference between star-forming galaxies, AGN, and composite objects (as defined by \citealt{Juneau11}).     
 While the evolution of MZ relation compromises somewhat the use of this method, we note that two of our galaxies are found in the area identified as possible AGN.  Since metallicity evolution works in the sense of shifting the proxy relation between  \ha/\nii\ $\lambda 6583$  and mass,  it shifts the star-forming locus (and black threshold curves) to the right in Figure \ref{bpt_mex}.   Therefore, the two objects in the AGN or composite part of the sample may in fact be part of the non-AGN distribution;  however, we take the conservative step of excluding them from further analysis.    
  We list the measured and derived properties of the two candidate AGN at the ends of Tables \ref{meas} and \ref{derived}.

\section{Calculating oxygen Abundances} 
\label{calc_metal}  
Interpreting metal abundance measurements requires that we account 
for systematic uncertainties. 
Different methods for measuring strong oxygen abundances yield 
results which are offset by up to 0.7 dex \citep{KE08, LS12, AM12}. 
On one hand, photoionization models that are used to derive theoretical calibrations (i.e.\ \citealt{KD02}) are often based on simplistic assumptions about \ion{H}{2}  region geometries and ionizing spectra 
\citep{vanZee}.    On the other hand, empirical methods, which correlate line ratios with \ion{H}{2} region electron
temperatures (i.e. \citealt{PP04, P05}), generally give lower metallicities than photoionization model-based calibrations \citep{McGaugh, KD02}.   Because electron temperature
measurements may be overestimated when temperature and density gradients are present in \ion{H}{2} regions, it is possible that the empirical calibrations are biased towards low metallicities \citep{PC69, Stasinska}.    
Alternatively, \cite{Nicholls} have suggested that discrepancies between electron-temperature measurements and theoretical strong-line estimates
 can be explained if the electrons in \ion{H}{2} regions deviate from an equilibrium Maxwell-Boltzmann
distribution.

Ultimately, the differences between metallicity calibrations are still not fully understood, but the systematic offsets can be accounted for using a set of transformation equations given by \cite{KE08}.  
In the sections that follow (except for \S 6), we use the calibration from Kobulnicky \& Kewley (2004; hereafter KK04), which is the 
mean of two theoretical R23 calibrations in the literature (\citealt{McGaugh} and \citealt{KD02}).       In \S 6 we instead use the calibration from 
\cite{Maiolino08} so that we may compare our data to results from \cite{Mannucci11}.

\subsection{High-Metallicity or Low?  Determining the Branch of R23} 
Metallicities derived from R23 can be degenerate, since this diagnostic is double valued.   At high metallicities, the oxygen lines are weaker because cooling is
efficient and the \ion{H}{2} regions have lower electron temperatures.  On the other hand, at low metallicities, the overall decrease in oxygen relative to hydrogen also
imprints a decrease in R23 with decreasing metallicity.  The ``turnover'' metallicity that demarcates the transition between the upper and lower branches of R23 depends on the ionization parameter\footnote{The ionization parameter is defined as the ratio of the ionizing photon density to the hydrogen density.  It can be written as $U = Q/4 \pi r^2 n_H c$, where $Q$ is the ionizing photon rate, $r$ is the radius of the \ion{H}{2} region, $n$ is the hydrogen density, and $c$ is the speed of light.  The parameterization $q = U \times c$ is also commonly found in the literature. }, and differs among the calibrations that are in the literature.   For the typical ionization parameters of our galaxies, and the KK04 calibration,  the turnover metallicity is around 12+log(O/H) $\sim 8.4$, which is reached when log(R23)$\sim 0.9-1.0$. 

Many different methods have been suggested for breaking the degeneracy between the upper and lower branch.  The most rigorous of these is to compare 
to an alternative metallicity diagnostic.  \cite{KE08} advocate for the use of the \nii\ $\lambda 6583$/\oii\   or \nii\ $\lambda 6583$ /\ha\ ratios to first provide a metallicity estimate that identifies the branch.  
However,  this method is more challenging at $z>0.5$ (requiring infrared spectroscopy to reach very faint   \nii\ $\lambda \lambda 6548, 6583$ lines), and impossible from the ground at $z> 2$.  The R23 diagnostic, on the other hand, can be measured from the ground out to $z\sim 3$ \citep{Maiolino08}.  It is therefore imperative that we learn how to break the R23-metallicity degeneracy without observations of the  \nii\ $\lambda \lambda 6548, 6583$ lines. Here, we examine other methods for establishing the R23-branch that have been proposed in the literature.

\begin{figure*} 
\plotone{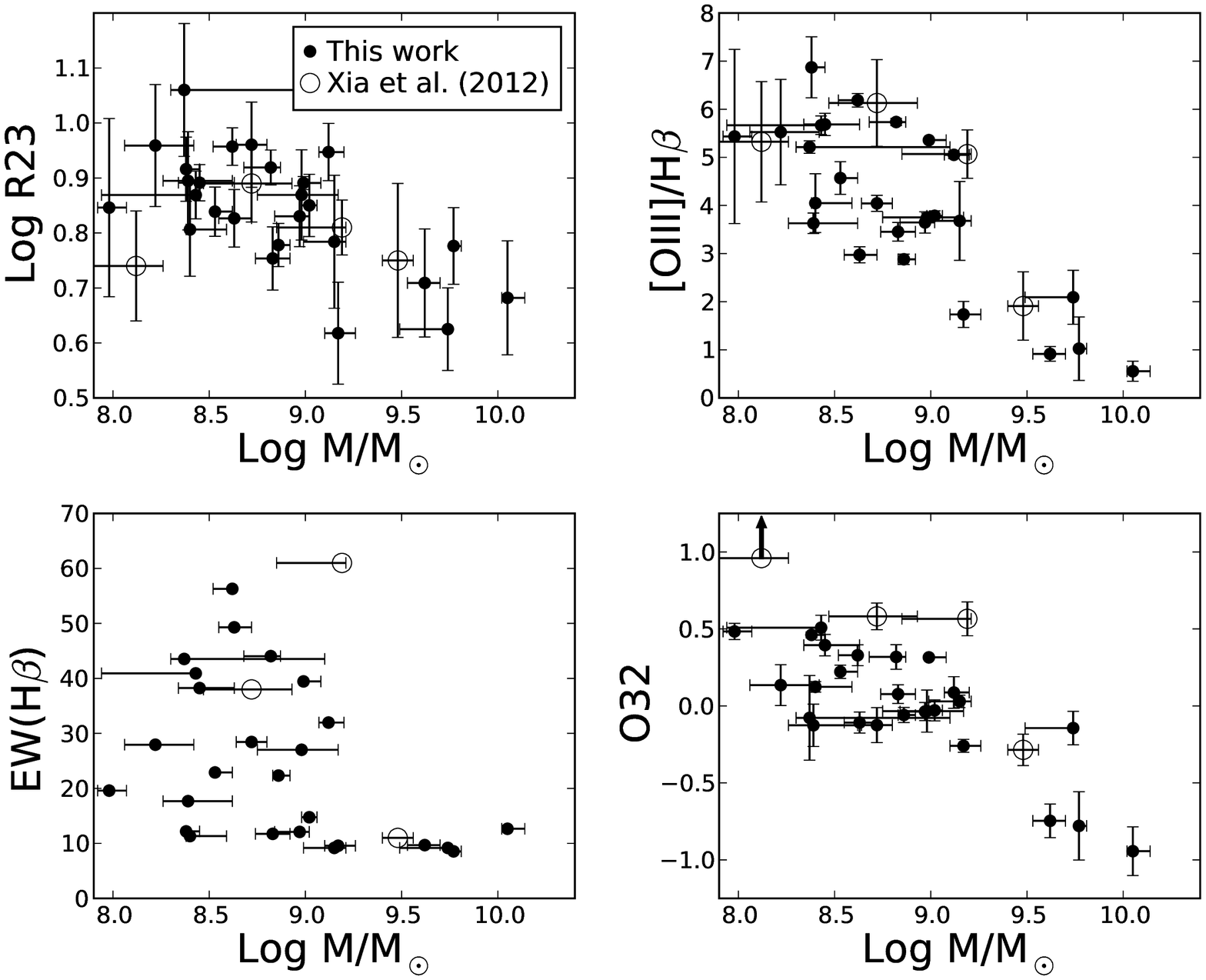} 
\caption{Our measured values of R23 decrease with increasing stellar mass, implying that most of the galaxies in our sample fall on the 
upper branch of R23.  This diagram is contrasted with three other diagnostics that have previously been used to identify whether galaxies have low or high metallicities \citep{Maiolino08, Hu09, Xia}.   These include the \oiii $\lambda \lambda 4959, 5007$ / \hb\ flux ratio, the 
rest-frame equivalent width of \hb, and the ratio O32 $\equiv$ log (\oiii $\lambda \lambda 4959, 5007$ /  \oii\ $\lambda \lambda 3726, 3729$). With an \hb\ EW of 352\AA\  (rest),  object 246 from Xia et al. is not shown in the  EW(\hb) panel.  }  
\label{r23vmass} 
\end{figure*}

In Figure \ref{r23vmass} we investigate the branch of R23 by plotting four different diagnostics as a function of 
stellar mass.   For comparison, we show the four \hb-detected galaxies ($0.60 < z < 0.85$) from the PEARS Survey (Probing Evolution and Reionization Spectroscopically; \citealt{Xia}).  In the upper left panel, we show that R23 decreases with increasing stellar mass for most of our sample.   Furthermore, three of the four galaxies from Xia et al. follow the trends from our emission line selected sample.  This diagnostic paints a clear picture:  if metallicity decreases as a function of decreasing mass, most of the galaxies in our sample must fall on the upper branch of the R23 indicator.  The two lowest mass galaxies shown in Figure \ref{r23vmass} (1217842 from our sample, and  246 from Xia et al.)  hint at a possible turnover at $M \sim 8.0-8.5$, indicating that these galaxies may fall on the lower branch of R23.  

Because it is not ideal to introduce a mass dependence on our metallicity derivations, we explore other emission line diagnostics in Figure \ref{r23vmass}.  First, we compare to the emission line ratios of \oiii/\hb\ and log(\oiii/ \oii) $\equiv$ O32 as a function of stellar mass.  These line ratios have been proposed to break the R23 degeneracy by  \cite{Maiolino08}, with low metallicity solutions being preferred when \oiii/\hb $>5$ and O32 $> 0.5$.     Indeed, Figure \ref{r23vmass} shows that the galaxies above these thresholds also prefer to have higher values of R23 (and lower metallicities on the upper branch).  Nevertheless, we still see very little evidence for a substantial population of galaxies extending to metallicities below the  R23  turnaround.     The only object that passes both the O32 and \oiii/\hb\ thresholds is object 246 from \cite{Xia}. 

The inference of upper branch metallicities for 3/4 of the objects presented by \cite{Xia} contrasts with their lower branch assumption.  
These authors argued for lower metallicities because of the relatively high values of \oiii/\hb\ and O32 for most of their sample.  However, comparison to the R23 vs mass plot (upper left) shows that for at least three of these galaxies, higher metallicity solutions are more plausible.  (Although the galaxies presented by \citealt{Xia} may prefer higher O32 and ionization parameters than the remainder of our sample, metallicity is not a strong function of ionization parameter on the upper branch, so their galaxies still follow our R23-stellar mass correlation.)   

Ultimately, caution is required when using  \oiii/\hb\ and O32 to determine the branch of R23, because these quantities  are dependent on the ionization parameter.   Observations of high-redshift galaxies show that their line ratios are offset to high values of \oiii/\hb\ at a fixed  \nii/\ha\ \citep{Shapley05,Erb06,Hainline09}.  This offset is usually interpreted as evidence for a systematically high ionization parameter  \citep{Brinchmann08}, suggesting that some galaxies deviate from the  local  metallicity vs. ionization parameter correlation.   In summary, attempting to break the R23 degeneracy using quantities that depend on the ionization parameter could erroneously indicate lower branch solutions. 

\begin{figure*} 
\plotone{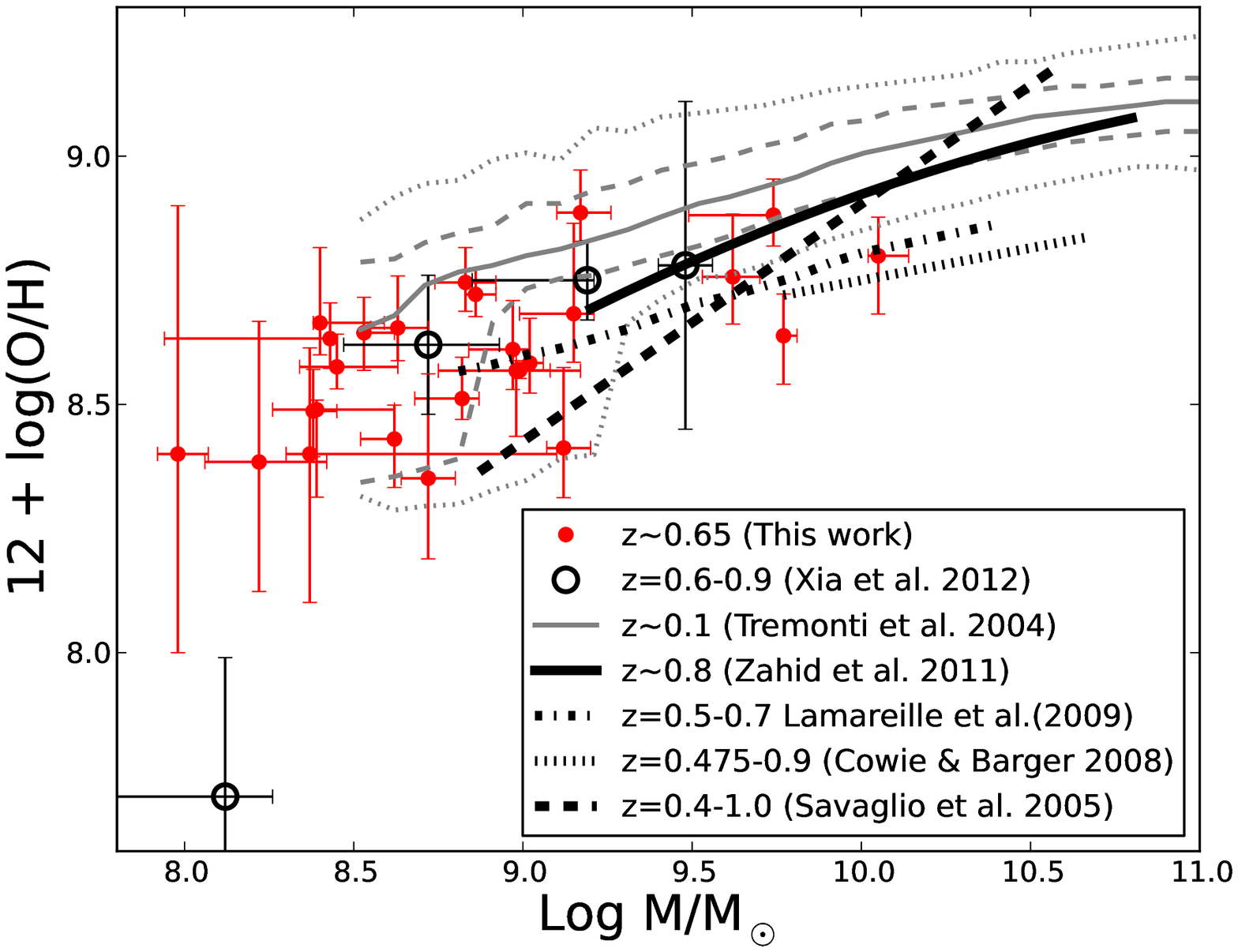} 
\caption{The addition of our 26 low mass galaxies constrains the low mass portion of the MZ relation at intermediate redshifts.  For comparison, we have re-calculated the metallicities presented in \cite{Xia}, assuming-- contrary to these authors--  that the three highest mass objects have metallicities on the upper branch.  Other studies are shown for comparison, and are (when necessary) converted to a \cite{Chabrier} initial mass function and a \cite{KK04} metallicity.  The dashed and dotted grey lines that follow the \cite{Tremonti} MZ relation show the 68 and 95\% contours for the SDSS data.  For reference, solar metallicity is 12 + log (O/H) = 8.69 \citep{solarox}.} 
\label{mz_compare} 
\end{figure*} 

In Figure \ref{r23vmass}, we also investigate the use of \hb\ equivalent width to select low-metallicity galaxies.    \cite{Kakazu} and \cite{Hu09} have found that, for their emission line selected sample, galaxies are likely to have very low metallicities (indicated by detectable \oiii\ $\lambda$4363 emission) when their \hb\ equivalent widths are more than 30 \AA\ (rest).     However, 8/26 galaxies in our sample meet this criterion, and we do not detect \oiii\ $\lambda 4363$ in any objects.  What is more apparent in Figure \ref{r23vmass}, is that a cut in \hb\ equivalent width merely selects against galaxies with $M \ga 10^{9.0}$ $M_{\sun}$.   As we show in \S \ref{mz_sec}, the Hu et al. sample has, on average, higher \hb\ equivalent widths
and somewhat lower luminosities than the present sample of emission line objects.    We infer that it is not straightforward to use \hb\ equivalent width to discriminate between upper and lower branch metallicities.

Guided by Figure \ref{r23vmass} we conclude that, in the absence of other metallicity diagnostics, the correlation between stellar mass and R23 is the preferred method to break the R23-degeneracy.   This exercise shows that for $z\sim 0.6-0.7$ galaxies, the maximum value of R23 is reached between $10^{8.0}$ and $10^{8.5}$ M$_{\sun}$.  Therefore, we adopt upper branch solutions above $M = 10^{8.2}$ $M_{\sun}$.    The two lowest mass galaxies  (1217842 from our sample and 246 from \citealt{Xia}) are less certain.   For 1217842 we adopt a metallicity in the turn around region (12 + log(O/H) = 8.4), with error bars denoting the upper bound of the high metallicity solution and the lower bound of the low metallicity solution. The lowest-mass object from Xia et al. (\#246) is tentatively assigned to the lower branch because of its more extreme line ratios, although  we caution that the \hb\ measurement could be in error because it is blended with \oiii\ in their low resolution data. Followup spectroscopy can
better constrain the metallicities of galaxies near the R23 turnaround by providing the \nii/\ha\ and \nii/\oii\ ratios \citep{KE08}.  
At present, however, the conclusions drawn in the remainder of this work do not depend on these two galaxies because their metallicity errors are large.   Oxygen abundances for the entire sample are listed in Table \ref{derived}.

Our inference of upper branch metallicities at $M>10^{8.2-8.5} M_{\sun}$ is inconsistent with results reported by \cite{Zahid11}.  In contrast to Figure \ref{r23vmass}, Zahid et al. find that the mean value of R23 turns over around $M\sim 10^{9.2}$ M$_{\sun}$ for their $z\sim 0.8$ sample.   
They interpret the turnover as evidence for lower branch metallicities.  However, as can be seen in their Figure 4, the data are increasingly noisy below $M < 10^{9.5}$ M$_{\sun}$,  and objects with R23 $ >  10$ have been removed as candidate AGN.   The coupling of these effects may bias the lowest mass bin  towards lower mean values of R23, and mimic the effect of a turnover in the R23 vs mass correlation.   We conclude that on average, galaxies at $z\sim 0.6-0.7$ fall on the upper branch of R23 for $M\ga10^{8.2-8.5} M_{\sun}$.  At higher redshifts, where metallicity evolution is significant  (i.e.\ \citealt{Erb06}), we expect  that the transition from the upper to lower branch of R23 will occur at higher stellar masses.

\begin{figure} 
\plotone{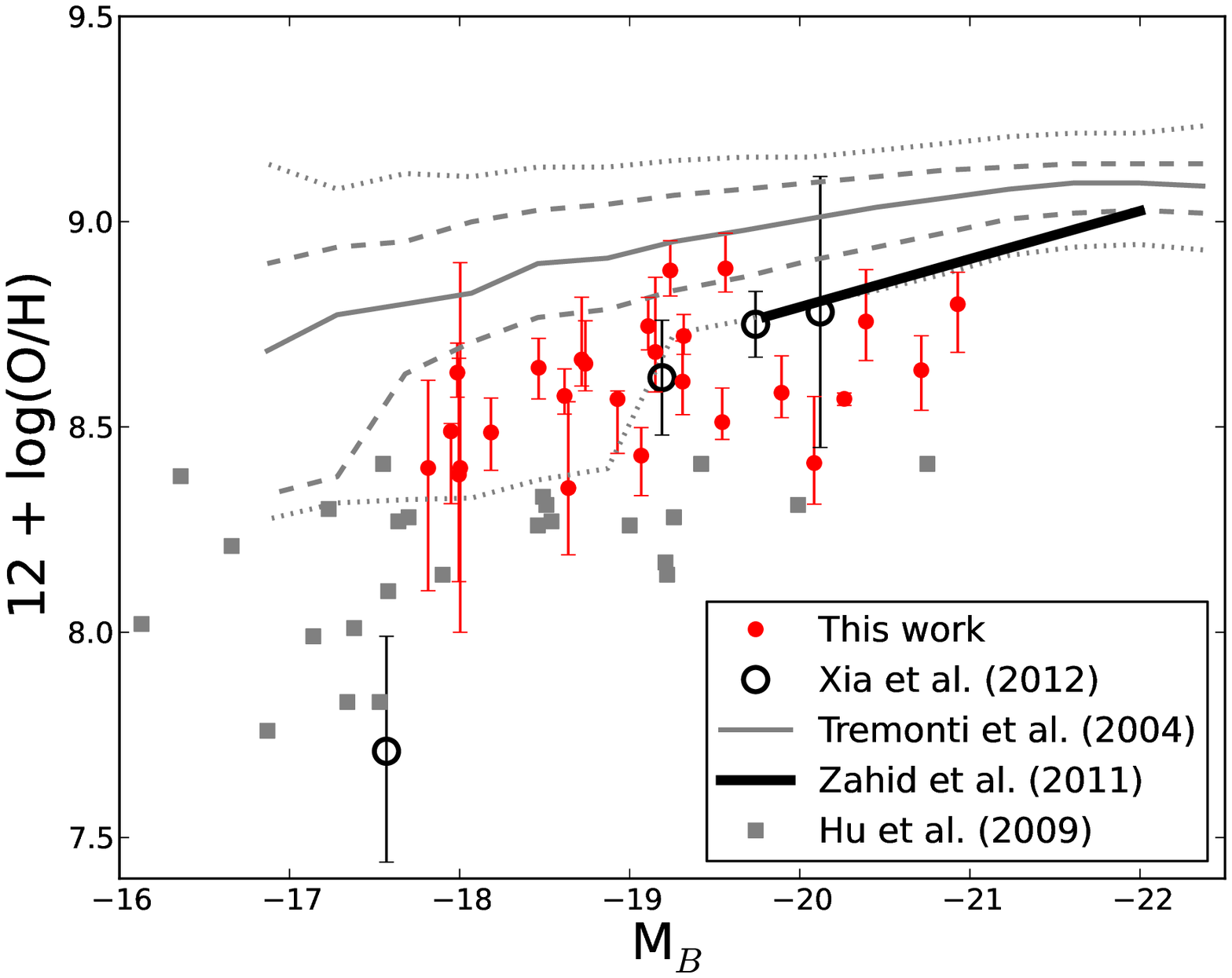}
 \caption{The luminosity metallicity relation for the present emission line selected sample is compared to correlations from the literature.  Notably, we compare to the measurements of the ultra-strong emission line galaxies reported by \cite{Hu09}. Because these 
 galaxies have detected \oiii\ $\lambda$ 4363 \AA\ emission, we assume that they fall on the lower branch of R23.  Their metallicities are recalculated using the KK04 calibration (assuming no dust, as justified by Hu et al.).     
  }
 \label{lzfig} 
 \end{figure}

\begin{deluxetable*} {crcccccccc}[!ht]
\tablecolumns{10}
\tablewidth{0pt}
\tablecaption{Emission Line Measurements} 
\tablehead{
\colhead{IMACS ID}  & \colhead{COSMOS ID} & \colhead{RA}  & \colhead{Dec} & \colhead{$z$} & \colhead{F(\oiii\ $\lambda$5007)} & \colhead{F(\oiii\  $\lambda$4959)}  & \colhead{F(\hb)} & \colhead{F(\oii\ $\lambda 3727$)} & \colhead{EW(\hb)}  \\
   &    &   \colhead{(J2000)} & \colhead{(J2000)} &   &  & &   &  & \colhead{\AA\ (rest)}  
  } 
\startdata
27.5+5-0.47 &	799604	   &  10:00:11.314 &  +02:04:07.60   & 0.675 &   $16.7\pm6.3$  &	   \nodata             &     $ 39.2 \pm 0.7 $  &	    $ 53.9 \pm 1.3$  & 12.7 \\ 
99.5-3-0.84    &	1235118	    &  10:00:53.842 &  +02:22:06.88  & 0.685 &  $19.75pm12.7$ &        \nodata	                 &  $ 25.1\pm 0.6$  &	           $ 58.0  \pm 1.2$ & 8.5  \\
90.5+6-0.63  &	1265343	    & 10:00:09.101  &   +02:19:52.02   &0.678 &  $10.2\pm2.7$ &          \nodata	               &     $ 6.3 \pm 0.3 $&	          $ 16.4 \pm 0.4$ &  9.2 \\
100.5-2-0.41 &	1234271	    &   10:00:50.551 &  +02:22:22.66   & 0.683    &  $13.4\pm2.2$ &	     \nodata 	      &     $ 19.0\pm 0.4$ &          $  48.5 \pm 0.9 $ & 9.7  \\ 
65.5-1-0.69 &	1003946	  & 10:00:47.710  & +02:13:36.83   & 0.620  &  $26.0\pm0.4$   &      \nodata               & 	   $   19.4\pm 3.0$   &      $ 40.9 \pm 0.7$  & 9.6 \\
 87.5+4-0.86 &	1242162	  &  10:00:18.502 &  +02:19:06.82   &  0.656 &  $15.4\pm0.7$  &	   $5.4\pm0.4$  &	     $ 5.4 \pm 1.2$   &      	 $ 19.4 \pm 0.8$  & 9.2  \\  
95.5+4-0.94 &	 1236769	  & 10:00:21.331 &  +02:21:05.22  &  0.667 &   $158.5\pm1.4$  &	   $52.5\pm 0.5$  &    $40.8 \pm 0.5 $  &        $ 112.3 \pm  1.0 $ & 32.0  \\
35.5+5-0.23 &	794540	   & 10:00:09.038  & +02:06:08.16    &  0.620 & $57.0\pm0.4$ &	   $ 19.1\pm 0.4$  &   $ 19.6 \pm 0.5 $  &   $ 70.5 \pm  1.0$  & 14.8  \\
\nodata       &	1015084	  & 10:00:41.863 &  +02:09:12.66  & 0.678 &        $309.7 \pm0.9$  &	$104.1\pm0.5$  &	      $75.1\pm 0.4 $ &   $200.3\pm 1.0$  & 39.5 \\
54.5+6-0.09 &	1036285	 & 10:00:01.553  &  +02:10:51.89 & 0.684 &     $53.3 \pm 1.0$  &      $17.2\pm0.6$  &    $18.5\pm 0.3$  &          $49.6 \pm 0.6 $   & 27.0 \\ 
25.5+7-0.08 &	801017	   &  09:59:58.039  &   +02:03:38.89  &   0.638 & $36.9\pm0.4$  &	   $12.8\pm0.4 $ &   $ 13.2 \pm 0.8 $  &	    $ 46.9  \pm 1.0$  & 12.1  \\ 
23.5+2-0.33 &	775548	   &  10:00:32.287 &  +02:03:09.25  &   0.670 &  $32.8\pm0.7$ &	   $12.2\pm1.3$  &     $ 14.8 \pm 0.5 $   &	    $44.8\pm 1.1$ & 22.3  \\ 
\nodata       &  	1035959	  &  09:59:59.347 &  +02:10:57.10 & 0.628 &  $35.2 \pm 0.3$ &	         $12.7\pm0.2$ &	      $13.2 \pm 0.7$  &	$ 34.7 \pm  0.6 $ & 11.7 \\ 
\nodata      &	982000	 & 10:01:07.140 &  +02:12:08.99   &  0.621 &  $235.1\pm0.6$  &	        $77.2\pm0.5$  &   $53.3 \pm 0.7 $   &	  $ 112.7 \pm 0.9 $  &  44.0 \\
\nodata       &	1013478	  & 10:00:18.701 &  +02:09:41.74  & 0.635   &   $34.9 \pm0.3$ &          $12.1\pm0.3$  &     $11.2\pm 0.5$  &	        $40.7 \pm 0.6$   &  28.4 \\
\nodata      &	980802	 & 10:01:07.135 & +02:12:38.22    &  0.621 & $27.3\pm0.4$  &        $9.2\pm0.6$   &	      $ 11.9 \pm 0.7 $  &	   $ 40.5 \pm 0.7$ & 49.3 \\
28.0+7.1\tablenotemark{a}     &	799190	   & 10:00:03.035  & +02:04:13.98   & 0.638 &   $167.5\pm0.6$  &           $57.3\pm0.4$  &    $  35.2 \pm 0.8$   &	     $91.3\pm 1.0$ & 56.3  \\ 
\nodata     &	770439	   &  10:00:42.334 &  +02:05:43.53   &  0.629 & $34.4\pm0.4$ &	      $11.5\pm0.4$  &   $9.8 \pm 0.7 $   &	     $23.9 \pm  0.8 $ &  22.9  \\
\nodata        & 1009808	  &  10:00:44.558  & +02:11:18.04    & 0.616 &  $96.9\pm0.4$  &	   $32.5\pm0.5$ &	      $22.2 \pm 0.9 $ &	  $45.2 \pm 0.8 $ & 38.0 \\ \
\nodata        &	1237667	    &  10:00:52.368 &  +02:20:58.18  & 0.640 &  $59.2\pm0.4$  &	      $21.2\pm0.4 $  &   $13.6 \pm 0.4 $ &	    $24.9  \pm 1.0 $ & 40.9 \\ 
\nodata     &	772773	   &   10:00:25.567 &  +02:04:44.36  &  0.616 &  $21.5\pm0.5$  &	      $8.2\pm 0.8 $ &   $ 6.9 \pm  1.0$     &	    $22.3 \pm 1.0 $ & 11.3 \\
48.5-4-0.66 &	989316	  & 10:01:08.537 &  +02:09:21.70  &  0.674  &   $13.8\pm0.5$   &      $4.7\pm1.0 $  &	      $4.96\pm 0.2$ &	          $16.1\pm 0.7 $  & 17.7 \\ 
54.5+4-0.18 &	1010994	& 10:00:16.097  &   +02:10:51.60 & 0.634&          $28.7 \pm  0.2$  &      $10.0\pm0.3$ &      $5.4\pm 0.5$ &	       $ 13.4 \pm 0.5 $  & 12.2 \\ 
47.5+3-0.85 &	1015516	& 10:00:27.792  &   +02:09:06.43 & 0.679 &       $34.1 \pm 0.3$ &      $12.4\pm0.2$  &     $8.5\pm 0.2$  &	       $17.7 \pm 0.6 $  & 43.5 \\
\nodata        &	1010583	  &10:00:21.408  & +02:11:00.43  & 0.639 &   $45.2 \pm 1.5$  &	$14.6\pm0.8$  &	      $10.6\pm 2.1$  &    $ 28.5 \pm 0.6$  & 27.9 \\  
\nodata        &	1217842	    &  10:00:54.658 &  +02:19:01.81 &  0.633 &  $21.4\pm0.3$ &	     $ 8.4\pm 0.4 $  &	$ 5.1\pm1.7$  &	    $ 9.8 \pm 0.8 $ & 19.6  \\
\cutinhead{Objects lacking clear optical counterparts in the COSMOS data} 
52.5+8-0.69 &	\nodata 	&  09:59:51.374  &  +02:10:23.28 &  0.678   & $15.6 \pm 0.8$ &      $3.7\pm0.9$  &      $2.1 \pm 0.1$ &	       $ 3.6 \pm 0.4 $    & \nodata \\ 
45.5+7-0.25 & \nodata	& 09:59:55.699  &   +02:08:38.51 &  0.678   & \nodata                    &      $2.2\pm 0.7$  &	      $2.6 \pm  0.4$  &      $ 9.6 \pm 1.0 $    & 6.5 \\ 
46.5-7-0.04 &   \nodata      & 10:01:25.152 &  +02:08:52.34   & 0.672 &  $6.2  \pm0.5$  &	         $2.5\pm1.5$ &      $2.4\pm 0.8$  &	           $5.4 \pm 0.7 $  & \nodata  \\
65.5-2-0.14 &	\nodata	 & 10:00:50.904 &  +02:13:36.52   & 0.624 & $14.1\pm0.3$  &          $5.0\pm0.9$   &	      $5.4\pm0.2 $ &	  $6.8 \pm 0.8 $  & \nodata \\
47.5+6-0.90 & 	 \nodata 	 & 10:00:07.145  & +02:09:07.81    &  0.639 & $14.4\pm0.4$  &	  $ 5.0\pm3.8 $ &	      $3.4\pm 0.8 $ &	    $< 9.5$  & \nodata  \\ 
87.5-1-0.06   &	 \nodata	   &   10:00:48.792  &  +02:19:06.54 & 0.620 &  $ 15.4\pm0.3$  &	    $  5.3\pm 1.6$  &  $  6.0\pm 1.7$   &           $ 1.9 \pm 1.3 $  & \nodata  \\ 
\cutinhead{Candidate AGN} 
55.5-3-0.60 &	984577	  & 10:01:01.126  & +02:11:07.85   & 0.623 &   $39.5\pm0.4$  &	   $13.7\pm 0.4 $  &   $16.4 \pm 0.5$ &   $	 51.6  \pm 0.7 $ & 9.9  \\ 
73.5+3-0.54 &	 998509	 & 10:00:25.046 &  +02:15:38.15  &  0.642 &  $38.7\pm0.5$  &	   $12.4\pm0.6 $  &  $48.4\pm 0.8 $   &	   $ 93.3 \pm 0.9 $ &  9.4
\enddata
\tablecomments{Emission line fluxes are in units of  10$^{-18}$ erg s$^{-1}$ cm$^{-2}$.   Equivalent widths and \hb\ fluxes are given as measured (without applying the 1\AA\ stellar absorption correction.)     Objects lacking an IMACS ID were selected strictly from the COSMOS narrowband (NB816) imaging, as we described in \S \ref{obs}.   The \oii\ flux is the sum of the $\lambda \lambda 3726, 3729$ doublet fluxes. In six cases, one of the lines in the \oiii\ doublet lines fell on an OH sky line; in this situation, metallicities are derived by assuming a flux ratio of F(\oiii\ $\lambda$5007)/ F(\oiii\ $\lambda$4959) $ \equiv3$.   }
\tablenotetext{a}{This galaxy was drawn from our earlier (wider and shallower) survey first presented in \cite{Martin08}.} 
\label{meas} 
\end{deluxetable*}

\begin{deluxetable*} {crcccrccrcc}[!ht]
\tablecolumns{10}
\tablecaption{Derived Properties} 
\tablehead{
\colhead{IMACS ID}  & \colhead{COSMOS ID} & \colhead{$M_B$}  & \colhead{$U-B$} & \colhead{log M$_*$} & \colhead{log SFR} & \colhead{$A_V$}  &  \colhead{log R23} & \colhead{log O32} & \colhead{12+log(O/H)} & \colhead{log $q$} 
  } 
\startdata
27.5+5-0.47 &	799604	   &  -20.93 &  0.52   &  $10.05^{+0.09}_{-0.03}$  &  $1.23_{-0.21}^{+0.34}$ &	  $1.80^{+0.13}_{-0.48}$  &  $0.68\pm0.10$   & $-0.94 \pm 0.16$ & $8.80_{-0.12}^{+0.08}$  &$7.00_{-0.13}^{+0.09} $ \\ [1.0ex] 
99.5-3-0.84    &	1235118	    &  -20.71 &  0.45     &  $9.77^{+0.04}_{-0.00}$&  $0.95_{-0.00}^{+0.20}$ &    $1.40^{+0.08}_{-0.19}$       & $0.77\pm0.07$   & $-0.78 \pm 0.22$ & $8.63_{-0.10}^{+0.08}$ & $7.05_{-0.10}^{+0.14}$ \\ [1.0ex] 
90.5+6-0.63  &	1265343	    & -19.24 &   0.61     & $9.74^{+0.00}_{-0.25}$  & $ -0.54_{-0.00}^{+0.12}$  &   $0.20^{+0.30}_{-0.05}$ 	  & $0.63\pm0.08$ & $-0.14 \pm 0.11$ &   $8.88_{-0.06}^{+0.07}$ &$7.65_{-0.07}^{+0.12}$  \\ [1.0ex]
100.5-2-0.41 &	1234271	    & -20.39   &  0.42      &  $9.62^{+0.08}_{-0.09}$ &	 $0.62_{-0.05}^{+0.38}$   &   $1.00^{+0.28}_{-0.31}$   & $0.71\pm0.10$  & $-0.75\pm 0.19$ &  $8.76_{-0.10}^{+0.13}$  & $7.11_{-0.10}^{+0.06}$ \\  [1.0ex]
65.5-1-0.69 &	1003946	  & -19.56  & 0.42    &  $9.17^{+0.09}_{-0.07}$     &   $-0.45_{-0.06}^{+0.44}$ &   $0.60^{+0.07}_{-0.21}$  & $0.62\pm0.09$  &    $-0.26\pm 0.05$ &      $8.89_{-0.06}^{+0.09}$ & $7.55_{-0.11}^{+0.09} $ \\ [1.0ex]
87.5+4-0.86 &	1242162	  &  -19.15  &  0.44   &  $9.15^{-0.06}_{-0.16}$   &  $-0.42_{-0.00}^{+0.07}$   &   $0.00^{+0.18}_{-0.00}$ &  $ 0.78\pm 0.12$   & $0.03 \pm 0.04$ &    $8.68_{-0.10}^{+0.18}$ &  $7.66_{-0.12}^{+0.14}$  \\[1.0ex]  
95.5+4-0.94 &	 1236769	  & -20.08   & 0.37    &    $9.12^{+0.08}_{-0.05}$ & $0.30_{-0.22}^{+0.59}$   &	 $0.60^{+0.27}_{-0.40}$  &  $0.85\pm0.05$   &  $0.09\pm 0.10$  &     $8.41_{-0.10}^{+0.16}$  & $7.57_{-0.12}^{+0.14}$  \\ [1.0ex] 
35.5+5-0.23 &	794540	   & -19.89   & 0.30    &  $9.02^{+0.04}_{-0.04}$ & $ 0.02_{-0.05}^{+0.22}$ & 	  $0.20^{+0.40}_{-0.07}$   &  $0.85\pm0.06$  & $-0.03 \pm 0.07$ &  $8.58_{-0.06}^{+0.09}$ & $7.56_{-0.12}^{+0.07}$  \\ [1.0ex] 
\nodata       &	1015084	  & -20.26 &  0.21     &  $8.99^{+0.09}_{-0.00}$    & $0.33_{-0.00}^{+0.15}$   &	 $0.00^{+0.09}_{-0.00}$     &  $0.89\pm0.01$  & $0.32 \pm 0.02$ &    $8.57_{-0.02}^{+0.01}$   & $7.83^{ +0.02}_{-0.03}$ \\ [1.0ex] 
54.5+6-0.09 &	1036285	 & -18.93 &  0.40      &     $8.98^{+0.19}_{-0.23}$  &  $0.08_{-0.46}^{+0.38} $   &  $0.60^{+0.44}_{-0.60}$  &   $0.87\pm0.08$ & $-0.03 \pm 0.14$ &   $8.57_{-0.13}^{+0.02}$   & $7.54^{+0.14}_{-0.05}$  \\  [1.0ex]
25.5+7-0.08 &	801017	   &  -19.31  &  0.37  &   $8.97^{+0.05}_{-0.03}$  & $-0.06_{-0.25}^{+0.00}$  &    $0.20^{+0.22}_{-0.20}$ &  $ 0.83\pm0.06$   & $ -0.04 \pm 0.06$ &  $8.61_{-0.08}^{+0.10}$   & $7.56_{-0.08}^{+0.09}$  \\ [1.0ex] 
23.5+2-0.33 &	775548	   &  -19.32 & 0.34    &    $8.86^{+0.06}_{-0.03}$ &  $-0.33_{-0.16}^{+0.18}$ &    $0.20^{+0.14}_{-0.20}$  & $ 0.78 \pm 0.04$&	$-0.06 \pm 0.05$ &    $8.72_{-0.04}^{+0.05}$ & $7.61_{-0.08}^{+0.09}$ \\  [1.0ex]
\nodata       &  	1035959	  &  -19.11 & 0.37     &  $8.83^{+0.09}_{-0.09}$    & $-0.36_{-0.18}^{+0.20}$ &	 $0.20^{+0.23}_{-0.20}$     & $0.75\pm0.06$  &  $0.08 \pm 0.06$ &	$8.75_{-0.06}^{+0.07}$  & $7.74_{-0.11}^{+0.10}$ \\ [1.0ex]  
\nodata      &	982000	 &   -19.54 &  0.25     & $8.82^{+0.05}_{-0.14}$   &  $0.43_{-0.48}^{+0.23}$    &   $0.40^{+0.24}_{-0.26}$   & $0.92\pm0.03$  &	$0.32 \pm 0.08$ &    $8.51_{-0.04}^{+0.08}$ & $7.80_{-0.10}^{+0.15}$  \\ [1.0ex]
\nodata       &	1013478	  & -18.64 & 0.36     &  $8.72^{+0.08}_{-0.08}$     &  $-0.28_{-0.15}^{+0.54}$  &   $0.60^{+0.46}_{-0.24}$    &  $0.96\pm0.08$  & $-0.12 \pm 0.11$&  $8.35_{-0.16}^{+0.21}$  & $7.40_{-0.05}^{+0.09}$ \\ [1.0ex]
\nodata      &	980802	 &   -18.74  & 0.37    &  $8.63^{+0.09}_{-0.08}$  &  $-0.56_{-0.50}^{+0.33}$   &    $0.20^{+0.27}_{-0.20}$    & $0.83\pm0.05$  & $-0.11 \pm 0.07$ &  $8.65_{-0.07}^{+0.10}$ & $7.53_{-0.07}^{+0.11}$  \\ [1.0ex]   
28.0+7.1    &	799190	   & -19.07  & 0.24     & $8.62^{+0.00}_{-0.10}$ &   $-0.04_{-0.20}^{+0.46}$  &    $0.20^{+0.43}_{-0.07}$  &   $0.96\pm0.03$   & $0.33 \pm 0.07$ &  $8.43_{-0.10}^{+0.07}$  & $7.77_{-0.13}^{+0.07}$  \\ [1.0ex] 
\nodata     &	770439	   &  -18.46  &  0.36    &  $8.53^{+0.10}_{-0.00}$ &  $-1.08_{-0.00}^{+0.46}$ &	 $0.20^{+0.11}_{-0.20}$   &  $0.84\pm0.04$   & $0.22 \pm 0.04$  & 	$8.64_{-0.08}^{+0.07}$ & $7.80_{-0.07}^{+0.06}$  \\ [1.0ex]
\nodata        & 1009808	  &  -18.62  & 0.28     & $8.45^{+0.18}_{-0.11}$   &  $-0.13_{-0.35}^{+0.18}$  & 	  $0.20^{+0.27}_{-0.20}$ &  $0.89\pm0.03$  &	$0.39 \pm 0.07$ &    $8.57_{-0.04}^{+0.07}$ & $7.91_{-0.07}^{+0.11}$  \\  [1.0ex]
 \nodata        &	1237667	    & -17.99 &  0.30      & $8.43^{+0.01}_{-0.49}$ & $-0.51_{-0.02}^{+0.76}$ &	$0.00^{+0.82}_{-0.00}$  &   $0.87\pm0.04$ &	$0.51 \pm 0.08$ &      $8.63_{-0.06}^{+0.07}$  &  $8.05_{-0.27}^{+0.13}$  \\ [1.0ex]  
\nodata     &	772773	   &  -18.72  & 0.28       &  $8.40^{+0.19}_{-0.02}$  &	$-0.61_{-0.02}^{+0.30}$ &    $0.00^{+0.15}_{-0.00}$  &   $0.81\pm0.08$ &  $0.12 \pm 0.04$ &   $8.66_{-0.06}^{+0.15}$ & $7.72_{-0.06}^{+0.08}$ \\ [1.0ex]
48.5-4-0.66 &	989316	  & -17.94 &  0.35     &   $8.39^{+0.23}_{-0.13}$    & $-0.21_{-0.52}^{+0.37}$   &  $0.60^{+0.40}_{-0.60}$   &   $0.89\pm0.09$ &	 $-0.13 \pm 0.14$  &  $8.49_{-0.18}^{+0.02}$  & $7.44_{-0.18}^{+0.03}$  \\ [1.0ex] 
54.5+4-0.18 &	1010994	& -18.18  &   0.40       &    $8.38^{+0.07}_{-0.00}$  &  $ -1.41^{+0.06}_{-0.01}$  & $0.00^{+0.04}_{-0.00}$ &   $0.92\pm0.06$ & $0.46 \pm 0.02$   &   $8.49_{-0.09}^{+0.08}$   & $7.91^{+0.05}_{-0.05}$   \\ [1.0ex]
47.5+3-0.85 &	1015516	&  -17.82 &  0.36       &    $8.37^{+0.73}_{-0.07}$  &  $ 0.46_{-1.11}^{+0.30} $  &  $1.60^{+0.30}_{-1.54}$&    $1.06\pm0.12$  &	 $-0.08 \pm 0.25$ &   $8.40_{-0.30}^{+0.21}$   & $7.36^{+0.22}_{-0.19}$ \\ [1.0ex]
\nodata        &	1010583	  & -18.00  & 0.26      &  $8.22^{+0.20}_{-0.16}$    & $-0.06_{-0.66}^{+0.28}$   &  $0.60^{+0.38}_{-0.56}$	 &  $0.96\pm0.11$  & $0.13 \pm 0.13$  &   $8.38_{-0.26}^{+0.28}$     & $7.60^{+0.15}_{-0.17}$  \\ [1.0ex]   
\nodata        &	1217842	    &  -18.00 & 0.20       &   $7.98^{+0.09}_{-0.06}$ & $-0.48_{-0.32}^{+0.17}$ &	 $0.00^{+0.22}_{-0.00}$   & $0.85\pm0.16$  &  $0.49 \pm 0.05$ &    $8.40^{+0.50}_{-0.40}$  & $8.05_{-0.27}^{+0.22}$  \\
\cutinhead{Objects with ambiguous counterparts in the COSMOS data} 
52.5+8-0.69 &	\nodata 	&  \nodata  &  \nodata    &  \nodata &  \nodata &   \nodata &      $1.03\pm0.04$ & $0.73 \pm 0.05$	&      \nodata     & \nodata \\ [1.0ex]
45.5+7-0.25 & \nodata	& \nodata  &  \nodata     & \nodata  & \nodata &   \nodata  &	   $0.74\pm0.12$  & $-0.04 \pm 0.13$  &     \nodata     & \nodata \\  [1.0ex]
46.5-7-0.04 &   \nodata      & \nodata  &  \nodata   &  \nodata  &	 \nodata &    \nodata   &     $0.77\pm0.16$  &  $0.21 \pm 0.10  $ &	        \nodata   & \nodata  \\ [1.0ex]
65.5-2-0.14 &	\nodata	 & \nodata &  \nodata   &  \nodata  &   \nodata &    \nodata   &	   $0.68\pm0.03$ & $0.45 \pm 0.06 $ &	   \nodata   & \nodata \\ [1.0ex]
47.5+6-0.90 & 	 \nodata 	 & \nodata  & \nodata    & \nodata  &	 \nodata &  \nodata  &	     $0.86\pm0.14$ &	$0.61 \pm 0.35$ &    \nodata   & \nodata  \\ [1.0ex]
87.5-1-0.06   &	 \nodata	   &   \nodata   &  \nodata  &   \nodata  & \nodata & \nodata  &   $0.57\pm0.13$ & $1.04 \pm 0.29$ &             \nodata & \nodata  \\
\cutinhead{Candidate AGN} 
55.5-3-0.60 &	984577	  & -19.78  & 0.62    &  $10.25^{+0.01}_{-0.00}$  &  $0.67_{-0.16}^{+0.00}$   &	  $1.20^{+0.10}_{-0.20}$   &   \nodata &	 \nodata  & \nodata &  \nodata \\ [1.0ex]
73.5+3-0.54 &	 998509	 & -21.23   &  0.54    &  $10.23^{+0.00}_{-0.06}$ & $ 0.62_{-0.08}^{+0.55}$ &	  $1.40^{+0.20}_{-0.12}$  &   \nodata   &	  \nodata  & \nodata  & \nodata  
\label{derived} 
\enddata
\tablecomments{These quantities are derived from COSMOS imaging and our measured line ratios. Magnitudes and colors are in AB units and are derived by integrating the best-fitting SED under the appropriate bandpasses.  
 Stellar masses and SFRs are given in solar units, and were derived using FAST \citep{FAST}. For this derivation we used a \cite{Chabrier} initial mass function and a \cite{Calzetti} extinction curve.   The line ratios, metallicity, and ionization parameter ($q$) are derived, taking into account a 1\AA\ (equivalent width) correction to the \hb\ fluxes, as well as the dust correction and its associated uncertainty.  
 Metallicities  and ionization parameters are calculated using the \cite{KK04} calibration.  A dust correction has not been applied to the objects have ambiguous optical counterparts in the COSMOS imaging survey.  } 
\end{deluxetable*}

\section{Results}  
\label{mz_sec}

\subsection{The Mass-Metallicity Relation} 
Figure \ref{mz_compare} presents the MZ relation for our 26 star-forming galaxies, and compares it to both the local relation 
\citep{Tremonti}, and other  intermediate-redshift measurements  from the literature.   In each case, we have taken care to (when necessary) convert stellar masses to a \cite{Chabrier} IMF (using multiplicative constants given in \citealt{Savaglio, CB08}), and convert metallicities to a KK04 calibration (using equations given in \citealt{KE08}).   
In the mass range  $8.5 \le M_{\sun} \le 9.0$, our data give a median value of 12 + log(O/H) = 8.63, with an RMS scatter of 0.12 dex.

The intermediate redshift MZ relation has been measured for higher mass galaxies by several authors  \citep{Savaglio, CB08, Lamareille,  Zahid11}.   Among the higher mass portion of our sample ($M > 10^{9.0} M_{\sun}$, where we can compare directly to previous results)  there is good agreement with all of the published MZ relations.   While the addition of our data does not distinguish  between these relations, we note that the Zahid et al. sample stands out as the largest  (1350 galaxies, as opposed to fewer than 100 galaxies in each of the others).    These authors discuss  the differences between 
 \cite{Savaglio}, \cite{CB08}, and \cite{Lamareille}.  They conclude that  sample selection effects and linear fits which are biased by outliers
 may account for the differences among these works.    While these effects are likely important, we also note that different dust correction methods may be significant.    On one hand, \cite{Zahid11} and \cite{Lamareille} calculate 
 R23 from equivalent widths, as they are less sensitive to dust extinction.  Although this method is subject to systematic uncertainties correlated with galaxy colors, \cite{Liang} show that this effect has a relatively weak impact on metallicity (-0.2 to 0.1 dex).  Cowie \& Barger, on the other hand, take the approach of determining dust extinction from SED fits. However, they do not assume a higher dust extinction for nebular compared to stellar light, as their lower redshift sample suggests that these quantities are consistent.  Comparatively, the dust correction adopted by \cite{Savaglio} is cruder; they use the same 
 extinction ($A_V = 2.1$) for all galaxies.  If this correction is accurate for the median of their sample, then at high (low) masses the dust correction will be underestimated (overestimated), and metallicities will be overestimated (underestimated).   Qualitatively, this effect 
 could explain their steeper MZ slope.   Additionally, it is worth noting that our data are inconsistent with an extrapolation of the Savaglio et al.\ relation. 
 In the sections that follow, we take the results from \cite{Zahid11} as the best measurement of 
the high mass, intermediate-redshift MZ relation. 

The addition of our low mass data allows for new constraints on the evolution of the MZ relation.  In Figure \ref{mz_compare} we compare to the local relation measured from the SDSS \citep{Tremonti}.  In the mass range  $10^{8.5} \le M/ M_{\sun} \le 10^{9.0}$,  we find a mean metallicity that is 0.12 dex lower than the local relation (at  $M = 10^{8.75} M_{\sun}$).   This trend confirms that metallicity evolution 
is relatively slow from intermediate to low redshifts.  Additionally, as our data are consistent with an extrapolation of the results from \cite{Zahid11}, we
agree with their finding that  metallicity  evolution is more significant at lower masses than at higher masses.    This trend is qualitatively consistent with downsizing in the later phases ($z \sim 1-2$ to today) of galaxy evolution.    After high mass galaxies have assembled most of their stars they have also made most of the metals that they will ever produce.  At the same time, these higher mass galaxies are less effective at removing metals through winds, and they are also less efficient at accreting gas \citep{Dekel}.    As a result, 
higher mass galaxies are less capable of changing their metallicities after they've assembled most of their stars.  
Under this scenario, we would expect lower mass galaxies to evolve onto the present MZ relation at later cosmic times.      Hence, extending observations to lower masses at all redshifts will provide essential insights into the physics of galaxy assembly and metal production.   We compare our data to theoretical predictions for the MZ relation in \S \ref{model_sec}.

\subsection{The Luminosity-Metallicity Relation} 
In Figure \ref{lzfig} we show the metallicity luminosity relation derived from our data.    While the MZ relation may represent a 
more fundamental correlation than the metallicity-luminosity relation, the latter allows comparison to studies where  
stellar-mass measurements have not been possible.   Here, we focus on the ultra-strong emission line (USEL) sample
presented  by \cite{Kakazu} and \cite{Hu09}.    Like the galaxies in our sample, the USELs  lie at intermediate redshifts and
are emission line selected.  However, the majority of the USELs are low metallicity galaxies with detected \oiii\ 4363 \AA\ emission. 
In Figure \ref{lzfig}, we compare the metallicity-luminosity relation of the USELs to our sample.  In order to make a fair comparison, 
we have calculated the metallicities of the USELs using R23 and KK04, assuming lower-branch solutions.  No correction for dust is applied, 
which \cite{Hu09} argue is appropriate for these faint, low mass objects.   It is worth noting that, under these assumptions, the majority of USELS have 12 + log (O/H) $> 8.0$, and would not be classified as extremely metal poor galaxies (12 + log(O/H) $< 7.65$ ; \citealt{Kniazev}).   However, this classification changes when direct method metallicities are used:  in this case, the oxygen abundances reported by \cite{Hu09}  are 0.03 - 0.74  dex lower  (0.4 dex on average).

Figure \ref{lzfig} shows clearly that the USELs have lower metallicities than the emission-line selected galaxies in the present work.
In fact, this trend is not surprising.  Because of the differing selections for the USELs and the present sample, the former have \hb\ equivalent widths that are two times larger (on average) than the latter.  This difference translates to higher specific (mass-normalized) SFRs for the USELs.  One possible explanation is that more actively star-forming galaxies will have B-band luminosities which are high for their stellar mass.   In this scenario, galaxies that fall along the MZ relation will move to the right when plotted in the luminosity-metallicity relation, as in Figure \ref{lzfig}.  However, another explanation is possible.    Within the local MZ relation, galaxies with higher specific SFRs have lower metallicities and follow a mass-metallicity-SFR plane called the Fundamental Metallicity Relation (FMR;   \citealt{LL10,  Mannucci10,  Mannucci11, Yates, AM12}).  Hence,  the differing metallicities between the USELs and the present sample supports the existence of the FMR in the relatively unexplored low mass, intermediate redshift regime.    We next  
provide the first quantitative test of the FMR for intermediate redshift galaxies below $10^{9} M_{\sun}$.

\section{The Fundamental Metallicity Relation} 
\label{fmr_sec}
In  the previous section, we showed that, for a fixed luminosity (and possibly mass), galaxies with higher 
equivalent width emission lines (i.\ e.\ the USELs) have lower metallicities.   This result supports the idea that 
star formation drives some of the scatter in the MZ relation, even out to intermediate redshifts.  
   For local galaxies at fixed stellar mass, objects with higher SFRs have lower metallicities than lower-SFR galaxies  \citep{Ellison08, Mannucci10, Mannucci11, Yates, AM12}.   \cite{Mannucci10} describe this relation as a plane, and refer  to it as the Fundamental 
 Metallicity Relation (FMR).    Remarkably, Mannucci et al. report that high-redshift galaxies also lie on the plane (albeit at lower masses and higher SFRs than most low-z galaxies), implying that the physics governing the evolution of metal enrichment has not changed over cosmic time.  This finding has recently been confirmed for a large sample of intermediate redshift ($z\sim 0.6$) galaxies from zCOSMOS \citep{Cresci}.  Nevertheless, tests for evolution of the FMR have not yet been extended below $M\sim10^{9}$ M$_{\sun}$.  Here, we provide the first such test.

In light of the significant differences between metallicities derived from various calibrations \citep{KE08}, it is essential that we use
metallicities which are derived consistently with those in (Mannucci et al. 2010, 2011; the latter work extends the initial analysis to lower masses). 
 These metallicities are based on the calibration from \cite{Maiolino08}, which is semi-empirical.    The high metallicities are derived from the photoionization models of \cite{KD02}, whereas  the lower metallicities are constrained from direct $T_e$ measurements reported by \cite{Nagao}.  While the Maiolino et al. calibration of R23 does not include a dependence on the ionization parameter, in other calibrations (i.e.\ KK04) this dependence is weak on the upper branch of R23, where the present sample lies.    Therefore, for this section, we re-calculate the metallicities for our sample using the \cite{Maiolino08} calibration of R23. 

\begin{figure} 
\plotone{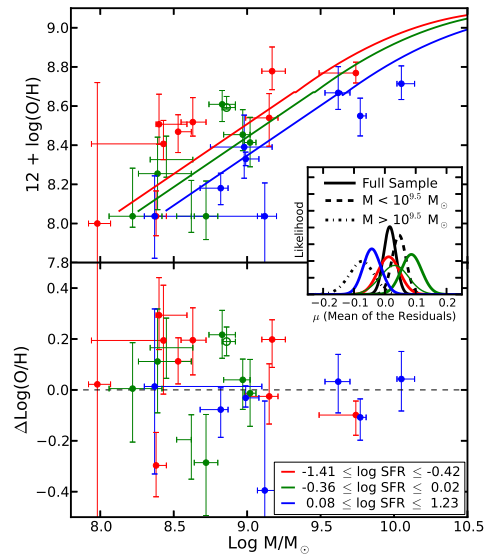} 
\caption{The fundamental metallicity relation, derived using the \cite{Maiolino08} metallicity calibration, shows that low mass galaxies with 
high SFRs have systematically lower metallicities.  The galaxy 23.5+2-0.33, which strongly influences the {\it weighted} mean of the residuals (see text), is shown as an open green symbol.   {\it Top-} The MZ relation is plotted, with galaxies color-coded in three bins of SFR.  
The curves show the local FMR reported by \cite{Mannucci11}.  FMR. {\it Bottom--} 
Residuals from the local FMR (data minus relation given in Equation \ref{fmr_eq}) show increased scatter towards low stellar masses, owing to larger measurement errors for faint galaxies.  
 (Note that residuals are calculated using measured SFRs rather than the mean values where Equation \ref{fmr_eq} has been evaluated for the top panel. As a result the residuals in the bottom panel do not exactly correspond to the differences between the data and models in the top panel.) 
{\it Inset--} Likelihood functions 
for the mean of the residuals are shown for different subsets of the data, including  the full sample,  low and high mass subsamples, and the SFR divided subsamples.   With one exception, the likelihood functions shown here include the galaxy 23.5+2-0.33, which drives some of the trends towards larger positive residuals (see text).    The thin green curve shows the effect of removing this outlier.  } 
\label{fmrfig} 
\end{figure}

Figure \ref{fmrfig} compares our galaxies to the $z=0.1$ FMR reported in \cite{Mannucci11}.   On the top panel, we show the MZ relation, with our data points  color coded by three different SFR bins.    The FMR, which is evaluated and shown for the same three SFR bins, is given by the relation (Equation 2 from \citealt{Mannucci11}): 
\begin{equation}
\begin{array}{ll}
12+{\rm log(O/H)}&=8.90+0.37m-0.14s-0.19m^2                    \\
           &+0.12ms-0.054s^2~~~~~~~~\rm{for}~\mu_{0.32}\ge9.5 \\
           \\
           &=8.93+0.51(\mu_{0.32}-10)~\rm{for}~\mu_{0.32}<9.5, \\
\end{array}
\label{fmr_eq} 
\end{equation}
where $m={\rm log}(M) - 10$,  $s = {\rm log}(\rm{SFR})$, and $\mu_{32} = {\rm log}(M) - 0.32 {\rm log}({\rm SFR})$.
This exercise shows clearly that a planar relation exists to lower masses at
$z\sim0.6-0.7$, with low SFR galaxies showing higher metallicities.  To highlight the differences from the FMR, the bottom panel shows the residuals from the local plane.  Here, we have included the uncertainty in the SED-derived SFR, as it enters into the metallicity that is predicted by the local FMR.  
 Overall, we see that most galaxies have metallicities that  fall within 0.2 dex of the local FMR, 
However, at low masses and low SFRs there are more objects with $\Delta$log (O/H) $>0$, so  it is reasonable to question whether these galaxies deviate slightly from the FMR.  Therefore,  we calculate the mean of the residuals, $\mu = \langle \Delta$log (O/H)$\rangle$, and its uncertainty by using a maximum likelihood estimation to account for the individual measurement errors.  Under this approach, the probability of observing a galaxy with a residual $y_i$ and error, $\sigma_i$ is  $p_i = e^{-(y_i - \mu)^2/2 \sigma_i^2 }$.   Then, for the full sample the likelihood function becomes  $\mathcal{L}(\mu)  = \prod p_i$.   We evaluate this likelihood function between $-0.4 < \mu < 0.4$, taking the maximum and the 68\% confidence intervals of  $\mathcal{L}(\mu)$ to represent the mean and its error. 
This estimate is made for the full sample, as well as low and high mass subsamples (divided at $10^{9.5} M_{\sun}$)  and the subsamples in three bins of SFR.  

 In the inset panel of Figure \ref{fmrfig}, we show the  likelihood functions for the mean of the FMR residuals.   
  These calculations indicate  only insignificant deviations from the local FMR.      In Table \ref{resid_table}, we give the mean of the residuals and its uncertainty for the full sample and various subsamples.  At first glance,  we see two weakly significant deviations from the local plane: one among the lower mass  ($M < 10^{9.5}$ M$_{\sun}$)  galaxies, and the other for 
those with intermediate SFRs (green points).  However, upon closer inspection we find that these trends arise solely because of one galaxy (23.5+2-0.33) with a metallicity  3.4$\sigma$ above the fundamental plane. (Furthermore, this galaxy has a strong influence on our calculation not because its residual is large, but rather because its errors are small.)  
Without this galaxy, the deviation disappears (see the thin green curve in the inset of Figure \ref{fmrfig}), and the mean of the residuals fall within 1.5$\sigma$  of zero for all of the subsamples listed in Table \ref{resid_table}.     Since any claims of a deviation from the FMR should be based on more than one object, we conclude  that our data show no compelling evidence for evolution at $10^{8} < M/M_{\sun} < 10^{9.5}$.   
  This result lends greater leverage to the finding that the FMR is not evolving for higher mass galaxies at these redshifts \citep{Cresci}.  
 
\begin{deluxetable} {l r} 
\tablecolumns{2} 
\tablecaption{Residuals from the local FMR} 
\tablehead{ \colhead{Sample}  & \colhead{Mean of the Residuals, $\mu$}  } 
\startdata  
Full sample                              &  $0.016 \pm 0.020$  \\
$M > 10^{9.5}$ M$_{\sun}$  &  $-0.074 \pm 0.083$ \\ 
$M < 10^{9.5}$ M$_{\sun}$  &  $0.046 \pm 0.022$ \\
$-1.41 \le$  log SFR $\le -0.42$ & $0.013 \pm 0.035$ \\ 
$-0.36 \le$ log SFR $\le~~ 0.02$ &  $0.086 \pm 0.033$ \\ 
~~$0.08 \le$ log SFR $\le~~ 1.23 $ &  $ -0.042 \pm 0.029$ \\ 
\cutinhead{Without 23.5+2-0.33}  
Full Sample   & $-0.009 \pm 0.021$ \\
$-0.36 \le$ log SFR $\le 0.02$ &  $0.027\pm 0.043$  \\  
$M < 10^{9.5}$ M$_{\sun}$ & $0.018 \pm 0.025$ 
\enddata 
\label{resid_table} 
\end{deluxetable}

It is also interesting to examine the scatter about the FMR, as intrinsic scatter could indicate that there are additional physical processes that  influence gas-phase metallicities.     
For intermediate redshift galaxies, \cite{Cresci} report a scatter of 0.11 to 0.14 dex about the FMR.  Yet, in our sample, the scatter appears to be larger 
at lower masses.    We find an RMS of 0.2 dex below  $M< 10^{9.5}$ M$_{\sun}$.   
Assigning some galaxies to the lower branch of R23 does not reduce this scatter.  The amplitude of the residuals are larger (or the about the same for a few galaxies) when the lower metallicity solutions are adopted.   In order to disentangle intrinsic scatter from increased measurement errors in faint galaxies, we use a Monte Carlo simulation. 
 First, we assume that there is no intrinsic scatter relative to the FMR.  Then, focusing only on the galaxies with $M< 10^{9.5}$ M$_{\sun}$, we generate 10,000 mock realizations of the FMR residuals, perturbing the data in Figure 7 by their uncertainties.     We calculate the RMS about the FMR for each realization, and find a mean of 0.23 dex, in good agreement with observations. This exercise implies that the amount of scatter that we observe can be entirely explained by our measurement errors.   Therefore, we conclude that the intrinsic scatter about the FMR must be small.  Repeating the simulation with additional intrinsic scatter shows allows us to place an upper limit on this scatter.  We find that  an intrinsic scatter of 0.16 dex produces an observed RMS $\le 0.2$ dex in only 5\% of Monte Carlo realizations.   Remarkably, the scatter about the FMR for the present sample is significantly smaller than the 0.4 dex RMS found for similar mass galaxies in the local universe \citep{Mannucci11}.  
We therefore conclude that the increased scatter in the low mass, $z=0.1$ FMR is likely the result of increased measurement uncertainties for faint galaxies.

\begin{figure*} 
\plotone{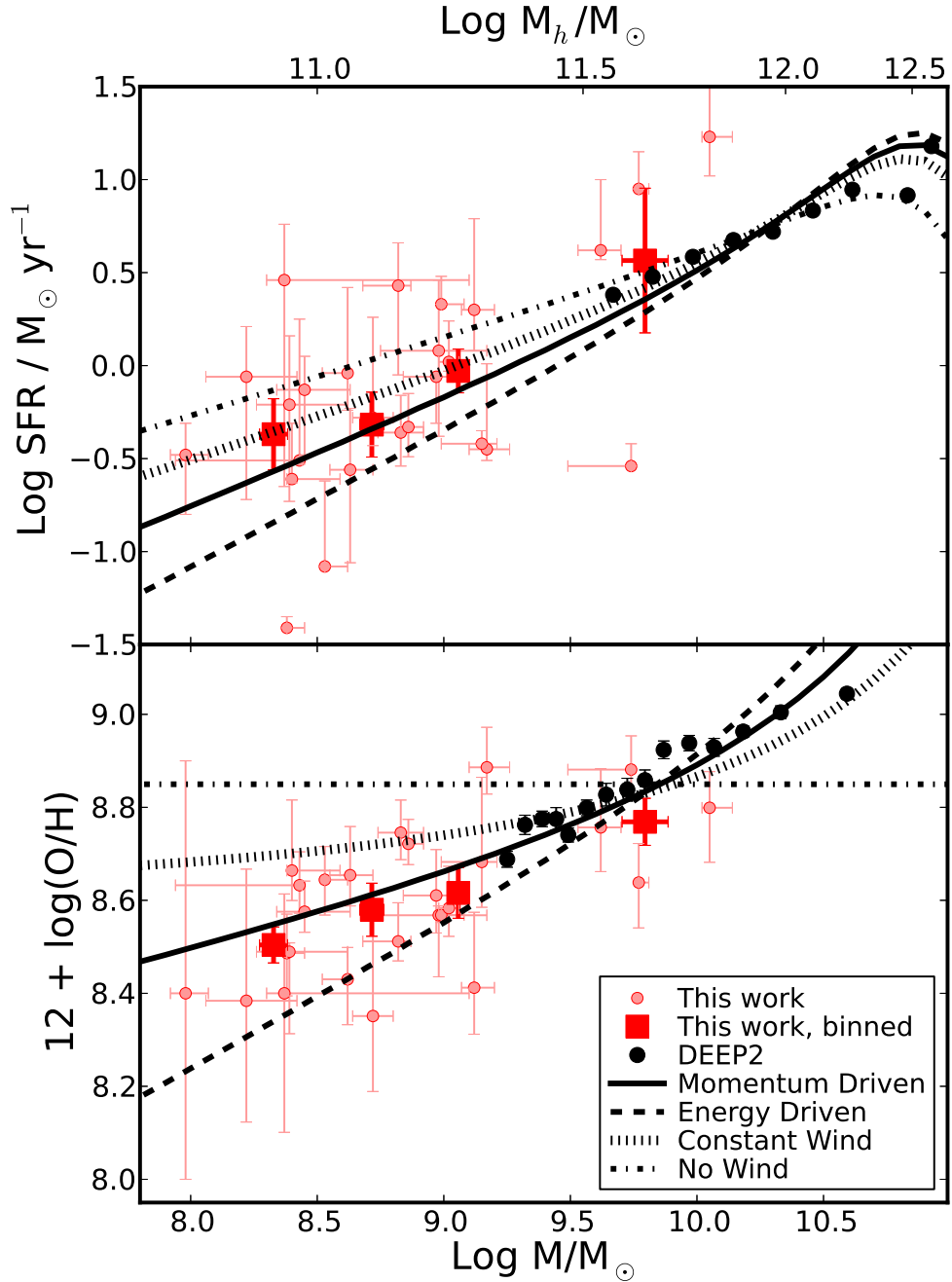}
\caption{The SFR-M$_{*}$ correlation  (top) and MZ relation (bottom) constrain models  from \cite{Dave12}.   Large red squares show the 
average mass, metallicity, and SFR for four mass bins: $M > 10^{9.5}  M_{\sun}$, $10^{8.9} <  M/M_{\sun} < 10^{9.2}$,  
$10^{8.5} <  M/M_{\sun}  < 10^{8.9}$ and $M < 10^{8.5} $ M$_{\sun}$.   Error bars on the binned data represent the standard deviation of the mean.  
We combine
our data with binned higher mass data from DEEP2:  the $z= 0.5-0.7$ SFR - M$_*$ correlation reported by Noeske et al. (2007; converted to a \citealt{Chabrier} IMF), and the $z\sim0.8$ MZ relation from \cite{Zahid11}.  MZ models are normalized by taking a nucleosynthetic yield that minimizes $\chi^2$ for the union of our data and the Zahid et al. data.  SFR-M$_*$ models are not renormalized.   The models are described in detail in \S \ref{model_theory}.  The halo mass scale shown at the top assumes the $M_* - M_{halo}$ relation from \cite{Moster}.      } 
\label{mz_models} 
\end{figure*}

\section{Constraining Gas Flows with the Mass-Metallicity and SFR-M$_{\star}$ Relations } 
\label{model_sec} 
As we introduced in \S 1, the MZ relation is an important measure of galaxy evolution.  Galaxies are not closed boxes; 
gas must be accreted from the IGM to fuel star formation (e.g.\  \citealt{Erb08,Bouche}), and galactic outflows are commonly observed at both low and high-redshift (e.g.\  \citealt{Heckman90, Martin05, Henry07, Soto12, Kornei12, Erb12}).    The MZ relation provides an independent, albeit indirect, constraint on these gaseous flows.   On one hand, 
the accretion of primordial gas will reduce the gas-phase metallicity in a galaxy.  On the other hand, galactic winds can create a deviation from
the closed-box model by lowering the gas fraction without changing the metallicity \citep{Edmunds, Dalcanton07}.
Combining our data with the higher mass MZ relation  provides an important increase in dynamic range that allows for more stringent constraints on models. 

Our emission line selection of galaxies does not strongly bias our sample or hamper our ability to test  theoretical models.    
In Figure \ref{mz_compare} we show that our sample is consistent with the DEEP2 MZ relation reported by Zahid et al.\ (2011; at least in the limited mass range where our data overlap). This agreement can be understood, since the FMR dependence 
on the SFR is relatively weak.  
For example, if the SFRs of our galaxies were biased high by a factor of several,  then (following the \citealt{Mannucci11} FMR) an ``unbiased'' sample would have metallicities which are approximately 0.1 dex higher.   An increase in metallicity this large is implausible, as it would imply a lack of metallicity evolution inconsistent with other studies.   Likewise, it is worth pointing out that the high S/N of the spectra shown in Figures \ref{allspec} and \ref{example_spec} implies that our sample is not biased towards higher metallicities by a requirement for strong \hb\ emission.  
The objects which were excluded due to low S/N all fell among 
the class \#3 objects discussed in \S 2.1;  these are either extremely faint, low-mass galaxies with uncertain optical counterparts, or brighter galaxies that fell 
partially outside our slit.  These galaxies have poorer quality spectra overall; it is unlikely that their exclusion biases our results.

In order to further verify that our sample is not biased by emission-line selection, we place our galaxies on the SFR-M$_{*}$ relation (i.e.\ the star-forming main sequence) in the top panel of  Figure \ref{mz_models}.  This comparison demonstrates that our galaxies do not have extreme SFRs for their stellar masses.  Rather, they lie on an extrapolation of the  relation reported by \cite{Noeske}.   Our sample could only be significantly biased if the  ``true'' SFR-M$_{*}$ relation turns over and becomes 
steeper at low stellar masses.  While this trend is predicted to occur when photoionization heating suppresses star formation in low mass halos, it is not 
expected to be important above $M_{halo} \ga 10^{9} M_{\sun}$ \citep{Okamoto}.   While beyond the scope of this paper, it is worth noting that we do not detect the flattening of the SFR-M$_{*}$ relation at low masses which is reported by \cite{Pirzkal12}.    

\subsection{Theoretical Framework} 
\label{model_theory}
One set of models which have recently gained traction are those that adopt an equilibrium between mass inflow ($\dot{M}_{in}$), outflow ($\dot{M}_{out}$), and 
gas consumption via star formation ($ \dot{M}_{*}$;  \citealt{Bouche, Dutton, Finlator08, Dave12}).    These models provide a relatively simple interpretation of the FMR as a manifestation of stochastic star formation.   When a galaxy accretes gas, its metallicity is diluted and its star formation increases.    The increased star formation serves to consume the extra gas and 
return the galaxy to its equilibrium.  Likewise, a pause in gas accretion will lower star formation rates  and increase gas-phase metallicities until accretion resumes.  These models not only predict  a MZ relation, but also the SFR-M$_{*}$ relation that we have shown in Figure \ref{mz_models}. above.    Therefore, in the remainder of this section we leverage the two relations to gain the best possible constraints on the models. 

 Quantitatively, we can express the equilibrium condition as (reproducing Equation 1 from \citealt{Dave12}): 
\begin{equation} 
\dot{M}_{in} = \dot{M}_{out} + \dot{M}_{*}.  
\label{eq_eq} 
\end{equation} 
Following this assumption, \cite{Finlator08} and  \cite{Dave12} show that SFRs and metallicities can be written as: 
\begin{equation} 
SFR = {{ \zeta \dot{M}_{grav}} \over { (1 + \eta) (1 - \alpha_z)}}
\label{sfr_eq} 
\end{equation} 
and
\begin{equation}
Z_{O} = {{y \over {1 + \eta} } { 1 \over {1 - \alpha_Z}}}.
\label{zmodel} 
\end{equation} 
The quantities in Equations \ref{sfr_eq} and \ref{zmodel} are defined as follows: 

\begin{enumerate} 

\item $Z_{O}$ is taken to represent the mass fraction of oxygen in the ISM.  This quantity is converted to  the units of 12 + log(O/H)  
by taking log($Z_{O}$) = log(O/H) + log( ${3 \over 4} \times M_{O} /M_{H}$  ).  $M_O$ and $M_H$ are the atomic masses of oxygen and hydrogen. 

\item  $\dot{M}_{grav}$  is the cosmological baryonic accretion rate, taken from \cite{Dekel}.  Because this quantity (as well as others below) 
are expressed in terms of halo mass, we adopt the $M_{*} - M_{halo}$ relation from \cite{Moster}.  

\item $\zeta$ is a quantity that represents preventive  (rather than ejective) feedback from gas heating.   There are multiple forms of
preventive feedback, and they combine multiplicatively.   
In the range of halo-masses  where our objects lie, the most important of these is gravitational heating from virial shock formation in accreted gas.  
Following \cite{Dave12}, we take the analytical form given by \cite{FG11}:  
\begin{equation} 
\zeta_{grav} \approx 0.47  { \left (  {1 + z} \over 4 \right )}^{0.38} { \left ( {M_{halo}} \over {10^{12}~ M_{\sun}} \right ) }^{-0.25}.
\label{zeta_grav} 
\end{equation} 
At high masses star formation quenching associated with supermassive black holes becomes important.  
While we do not aim to constrain quenching, we implement it following \cite{Dave12}:
\begin{equation} 
\zeta_{quench} = \left ( 1 + 1/3 \left ({M_{halo}} \over {10^{12.3} M_{\sun} } \right ) ^2 \right ) ^{-1.5}.
\label{zeta_quench} 
\end{equation}
The mass scale that we have adopted for $\zeta_{quench}$ is for $z=0$, and while it may be higher at earlier cosmic times \citep{Dekel}, the exact scaling is unimportant in the present analysis. 
At the low mass end, we take the photoionization feedback to be unimportant; as mentioned above, it is not expected to play a significant role  
at $M_{halo} \ga 10^{9} M_{\sun}$. 
 Finally, \cite{Dave12} do note that heating by galactic winds may be important, but the physics of this effect is poorly understood.  As such, they adopt an arbitrary parameterization of $\zeta_{winds}$ to consider its qualitative effects.  As we will show below, this additional heating is not needed to reproduce the present data.

\item $y$  is the nucleosynthetic yield of oxygen, which is between $0.008 < y < 0.021$ \citep{Finlator08}.  Since metallicity
calibrations are uncertain \citep{KE08}, we take $y$ as a free parameter as we search for a normalization that best matches our observed MZ relation. Hence, our comparison to the MZ relation concerns its slope, but not the normalization. 
\item $\eta$ is the mass loading factor, which is defined as the ratio of the outflow rate to the SFR:  $\eta = \dot{M}_{out} / \dot{M}_{*}$.   
In the comparison that follows, we adopt mass-loading factors that  
correspond (respectively) to momentum- and energy- driven winds  (i.e. \citealt{Murray, Finlator08, Dave11b, PS11, Dave12}):
\begin{equation} 
\eta = (M_{halo}/10^{12} M_{\sun})^{-1/3}
\label{vzw} 
\end{equation} 
\begin{equation} 
\eta = (M_{halo}/10^{12} M_{\sun})^{-2/3}.   
\label{ew} 
\end{equation}  
The normalizations of these $\eta$ are taken from \cite{Dave12}.  In addition to these two parameterizations of $\eta$, we compare to the case of no winds  ($\eta = 0$),  also $\eta=1$, where the mass loss rate is fixed to the SFR.
\item  Finally, the quantity $\alpha_Z$ is the ratio of the metallicities of infalling gas and the ISM: $Z_{in}/Z_{ISM}$.  The inclusion of $\alpha_Z$ allows for enriched inflows (possibly from previously ejected material now being re-accreted from the galaxy halo).  Additionally, it appears in the  model for SFR as a simple way to quantify the recycling of enriched halo gas into the ISM (i.e.\ an additional source of fuel for star formation).      The parameterization of $\alpha_z$ is taken from momentum-driven wind  simulations in \cite{Dave11b, Dave12}: 
\begin{equation} 
\alpha_Z = (0.5 - 0.1 z) (M_{*}/10^{10} M_{\sun})^{0.25}   
\label{eq_az} 
\end{equation} 
(although they note that it is a ``crude'' parameterization). While we aim to test models beyond those that are momentum-driven, $\alpha_z$ is not reported for these simulations.      Therefore we use this parameterization for all of  the models with winds.  For the no wind case we consider $\alpha_z = 0$, since there can be no recycling of previously ejected material in this scenario.  
We consider the effects of modifying $\alpha_z$ in the discussion below. 
\end{enumerate}

  \subsection{Comparison of equilibrium models to intermediate-redshift data} 
Figure \ref{mz_models} shows how the equilibrium models compare to the intermediate-redshift MZ  and SFR-M$_{*}$ relations.  
In order to get the best constraints on the models,  we combine our sample with binned DEEP2 data representing the MZ relation
for $z\sim 0.8$ galaxies  \citep{Zahid11} and the SFR-M$_{*}$ relation for $z\sim 0.5-0.7$ galaxies \citep{Noeske}. 
As we mentioned above, since  metallicity calibrations and nucleosynthetic yields are uncertain, we fit for the yield that reproduces the MZ data.  
The yields that minimize $\chi^2$ (including uncertainties in both metallicity and mass\footnote{We take $\chi_i^2 = (OH_i - f(M_i))^2 / (\sigma_{OH, i}^2 + f'(M_i)^2 \sigma_{M, i}^2$), where 
$f(M_i)$ and $f'(M_i)$ are the model and its derivative evaluated at $M_i$.  $\chi^2$ is the sum over the $i$ observations of $\chi_i^2$.  }) are $y =  0.012, 0.013, 0.010$ and 0.008 for  the momentum driven, energy driven, constant wind ($\eta = 1$) and no wind ($\eta =0$, $\alpha_z = 0$) models respectively.    We next describe how each model compares to the data:

\paragraph{No wind} 
Not surprisingly, equilibrium models that exclude outflows are unable to reproduce the data.   Most drastically, when $\eta = 0$ and
$\alpha_z = 0$, Equation \ref{zmodel} has no mass dependence.   Hence, the predicted MZ relation is implausibly flat,  and the slope of SFR-M$_{*}$ relation 
is also too flat. Figure \ref{mz_models} shows that, without winds low mass galaxies make too many stars (and metals), and are unable to get rid of the metals that they make.     It is worth noting that $\alpha_z$ is not strictly zero in the no-wind simulation \citep{Dave11a}, since infalling gas can be enriched by other galaxies or could represent the accretion of a satellite galaxy.   A non-zero $\alpha_z$ would further increase the SFR and amplify the disagreement 
between the SFR-M$_{*}$ data and the no-wind model.

\paragraph{Constant Wind} 
A model where the mass-loss rate is equivalent to the SFR shows good agreement with the SFR-M$_{*}$ relation at all stellar masses.    However, this model  predicts an MZ relation which is too flat at low stellar masses.

\paragraph{Momentum Driven Wind} 
Of the models shown in Figure \ref{mz_models}, the momentum driven wind shows the best match to the MZ data.  However, the 
metallicities predicted by this model turn up  at higher masses, owing to the dependence on $\alpha_z$. 
Equation \ref{zmodel} shows that the predicted metallicities diverge as $\alpha_z$ approaches unity at higher
masses.   While the disagreement between the model and data is only apparent in the highest mass bin, the upturn remains
implausible.  Locally, the MZ relation flattens to higher masses \citep{Tremonti}, so the model shown in Figure \ref{mz_models}  crosses the $z\sim 0.1$ relation 
at $M \sim 10^{10.5}$ M$_{\sun}$.   This observation  suggests that the mass scaling for $\alpha_Z$ given by \cite{Dave12} may be too steep at high masses. 

In addition to the MZ relation, we compare the momentum-driven wind model to the SFR-M$_{*}$ relation in the top panel of Figure \ref{mz_models}. 
In this case, the model  slightly under-predicts the SFRs for the galaxies in 
the present study as well as  for the lower mass DEEP2 measurements.
This discrepancy between the data and models at low masses
is qualitatively similar to what is found at higher masses and lower redshifts.  
\cite{Dave11a} and \cite{Weinmann} find that both simulations and semi-analytic models
are unable to reproduce the population of low mass galaxies with high specific SFRs.  

Given the uncertainty in $\alpha_z$, we next consider whether modifying this variable can 
produce  a momentum driven wind model that is a good fit to both the MZ and SFR-M$_{*}$ data. 
If we remove the mass-dependence of $\alpha_z$ in Equation 
\ref{eq_az}  ($\alpha_z \propto M_{*}^{0.0}$), the predicted SFR-M$_{*}$ relation appears very similar to the constant wind model 
(a good match to the data), but the predicted MZ relation (not shown) is far too flat.    In summary, 
modifying $\alpha_z$ does not improve the agreement between the data and the momentum driven wind model.

\paragraph{Energy Driven Wind} 
The energy driven wind models shown in Figure \ref{mz_models} are too steep compared to both the MZ data and the SFR-M$_{*}$ data.   However, as we already noted, some of this steepening of the MZ relation occurs because $\alpha_z$ may be too large at high masses.  In fact,  an energy driven wind model with $\alpha_z \propto M_{*}^{0.0}$ and $y=0.022$  appears very similar to the momentum driven wind model shown in {\it both}
the top and bottom panels.  For this model, the MZ relation model reduces to a straight line that 
runs 
through all of the data, and SFRs are slightly under-predicted at low masses.   Ultimately, we cannot distinguish the momentum driven model from
an energy driven  model with  $\alpha_z \propto M_{*}^{0.0}$.  However, we do note that this variable shows some mass dependence in the 
momentum-driven wind simulations \citep{Dave11a, Dave11b}, so it unlikely that this dependence is non-existent for energy-driven winds.   
We therefore conclude that energy-driven winds are unlikely to explain the data.  

 Recent simulations from Hopkins et al. (2012)  have suggested that 
 outflows depend strongly on gas surface density, so that winds from low  
 density dwarf galaxies are dominated by energy-driving from supernovae and stellar winds.   
 More massive star-forming galaxies with higher gas densities, on the other hand, are predicted to prefer momentum driven
 winds.     The present data do not indicate a mass-dependent $\eta$; Figure \ref{mz_models}  shows no evidence for steeper MZ and  SFR-M$_*$  slopes towards lower masses. 
 
\vspace{0.1in} 
This analysis (and Figure \ref{mz_models})  shows that there may be a tension between the data and the equilibrium model.  
On one hand,  models which best match the MZ relation under-predict the amount of star formation 
in low mass galaxies  (and even intermediate masses probed by DEEP2).  Likewise, models that best reproduce the SFR-M$_{*}$ relation over-predict the metallicities of galaxies. 
This trend can be understood as a generic property of the models,  because both Equations \ref{sfr_eq} and \ref{zmodel} are  proportional to $(1+\eta)^{-1} (1 - \alpha_z)^{-1}$; 
 changes in these model components alter the MZ and SFR-M$_*$ relation {\it in the same manner}. 
 However, if preventive feedback from gas heating has been overestimated 
  ($\zeta$ underestimated in Equations \ref{sfr_eq}  and \ref{zeta_grav}), then  SFRs
  can be increased without modifying the MZ relation. 
 In fact, \cite{Dave12} note that  $\zeta_{grav}$ is taken from simulations  that do not include 
 metal-line cooling \citep{FG11}, and raise the question of whether this effect is important. 
 Alternatively, if  $\zeta$ is correct, then the data may suggest a deviation from the equilibrium model.

 Ultimately,  we conclude that this  disagreement remains subtle. While we have argued that emission line selection 
 does not largely bias our results, we cannot rule out a small effect. As an example, if the SFRs presented in Figure \ref{mz_models} are 
high by 0.1 dex, we would expect an unbiased sample to have an SFR-$M_*$ relation that agrees better with the momentum driven wind model at low masses.   Additionally,  
according to the FMR, SFRs which are 0.1 dex lower should be accompanied by metallicities which are around 0.03 dex higher, in even 
better agreement with the momentum driven wind MZ relation.     Even without emission-line selection bias, it is important to
recall that the [\ion{O}{2}], \hb, and SED-based SFR indicators discussed in the Appendix differ systematically by up to 0.2 dex.  
However, modifying the SFR under these scenarios can not satisfactorily bring the data in line with the momentum-driven model. 
Even in the higher mass data alone, this model predicts a steeper SFR-$M_{*}$  relation than the  DEEP2 result \citep{Noeske}; therefore, the discrepancy which we have identified is apparent in a magnitude-limited sample.  We conclude that \ha\ followup spectroscopy 
 of our sample (combined with a  mass selected control sample) can definitively assess the effects of
 emission line selection, while simultaneously reducing errors on dust, SFR, and metallicity. 
 We leave this work as the subject of a future paper.

Finally, it is worth pointing out that systematic uncertainties in our metallicity calibration cannot  explain the discrepancy between the
data and models. \cite{KE08} show that local (SDSS) MZ relations have similar low mass slopes under most calibrations.   The \cite{Tremonti} calibration  and the direct-method electron temperature calibration  \citep{AM12} are two exceptions-- both produce MZ relations that are steeper than what we have used here.  Hence, 
if we calculated metallicities in the same way as Tremonti et al. or by using electron temperatures, the tension between the data and models would be amplified.     On the other hand, the calibrations reported by \cite{P01} and \cite{P05}  give
MZ relations that are flatter than most.  However, we note that they also have metallicities around 0.5-0.7 dex  lower than those from KK04.  These lower metallicities would introduce other difficulties, as they would require normalizing yields that 
fall below the plausible range indicated in \cite{Finlator08}.

 \section{Conclusions}
In this work we have placed the first constraints on the low mass, intermediate redshift MZ
 relation.  By using emission line selection in the COSMOS field, we are able to efficiently identify and measure the metallicities from 26 galaxies reaching  masses of $10^{8}$ M$_{\sun}$ at $z\sim0.65$.   Combined with previous measurements from magnitude limited samples, these data extend our knowledge of the MZ relation to masses an order of magnitude smaller.    This limit is comparable to the low mass limit of the local MZ relation determined from SDSS data. Therefore, for the first time we are able to measure the metallicity evolution of intermediate redshift galaxies at $M < 10^{9}$ M$_{\sun}$.   Compared to the $z\sim 0.1$ relation reported by \cite{Tremonti}, we find an average metallicity that is 0.12 dex lower for galaxies with $10^{8.5} < M/M_{\sun} < 10^{9.0}.$  We show that this measurement is consistent with an MZ relation that evolves more strongly at low stellar masses.  
 We interpret this mass-dependent evolution as consistent with downsizing trends, where higher mass objects have less leverage to alter their gas-phase metallicities after most of their stars have been assembled.

An important development in our understanding of metallicity evolution is the discovery that the scatter in the relation
can be reduced by accounting for star formation.  This measurement was quantified as a plane, and dubbed the Fundamental Metallicity Relation (FMR) by \cite{Mannucci10}.   Using the present sample, we find that a planar relation
does indeed exist among low mass, intermediate redshift galaxies.   Consistent with the findings of Mannucci et al.,  we see no evidence for evolution of the FMR, as our emission-line selected sample falls in good agreement with their $z\sim0.1$ FMR \citep{Mannucci11}.  Furthermore,  while the lowest mass galaxies in our sample show significant scatter with respect to the FMR, we attribute this  scatter to measurement uncertainties for faint galaxies.   Using a Monte Carlo simulation, we
determine that the intrinsic scatter is consistent with zero, and should be less than 0.16 dex (95\% confidence) at $M < 10^{9.5}$ M$_{\sun}$.

The MZ relation is an important probe of galaxy evolution models, as accretion and galactic outflows  modulate the gas-phase metallicities of galaxies.  Hence, comparison to model predictions can help us to gain insight into galaxy formation.
We have combined our MZ relation with the SFR-M$_{*}$ correlation that is also 
measured from our data,  taking higher mass data from the literature \citep{Noeske, Zahid11}.  We compare to the family of models 
outlined in \cite{Dave12}, where it is assumed that galaxies  prefer to maintain an equilibrium between inflows, outflows, and star formation \citep{Finlator08, Bouche, Dutton, Dave12}.     We find that models which predict the MZ relation may under-predict the SFRs of low mass galaxies, and at the same time, models that predict the SFRs of low mass galaxies tend to over-predict their metallicities.  
While this finding could represent a breakdown of the equilibrium model in low mass galaxies, it could alternatively be an indication that feedback from gas-heating has been overestimated in simulations  (i.e.\ \citealt{FG11}).    To solidify this result, we have begun   \ha\ and \nii\  spectroscopy
to followup the present sample and simultaneously measure a mass-selected control sample. These data will clarify the effects of emission line 
selection and greatly reduce the statistical uncertainties in our metallicity, dust, and SFR measurements. 

To conclude, these observations have provided a valuable look at the metallicity evolution of low mass galaxies outside the local universe.   We look forward to the physical insights and new constraints that can be gained from larger samples and a better characterization of systematic uncertainties.

\acknowledgements
The authors thank Jane Rigby, Dawn Erb, Joey Wong, Amber Straughn, Evan Skillman, Molly Peeples, Nicolas Bouch\'e, Brian Siana and Susan Kassin
for insightful discussions.   We also wish to thank the anonymous referee for helping to improve this manuscript.  We are grateful to Esther Hu and Jabran Zahid for providing tabular data, and Peter Capak and the 
COSMOS team for the high-level science products that made this project possible. 
This research has made use of the NASA/IPAC Infrared Science Archive, which is operated by the Jet Propulsion Laboratory, California Institute of Technology, under contract with the National Aeronautics and Space Administration. 
 This work was supported by NSF grants AST-0808161 and AST-1109288.  
 The authors recognize and acknowledge the very significant cultural role and reverence that the summit of Mauna Kea has always had within the indigenous Hawaiian community. We are most fortunate to have the opportunity to conduct observations from this mountain.

\appendix 

\section{Assessing SFR measurements}  
In \S \ref{fmr_sec} and \S \ref{model_sec} we draw conclusions from SFRs that we derived from SED fits;  therefore, it is important that we use reliable SFRs.   In order to 
assess systematic uncertainties, we compare this measurement to two additional SFR indicators:  \hb\ and \oii.    For the \hb\ method, we use the dust extinction from 
the SED fits, scaled up by a factor of 2.3 to account for the difference between stellar and nebular extinction \citep{Calzetti}.   We also correct the \hb\ fluxes for the small 
amount of stellar absorption, as discussed above.    Next,  the \hb\ luminosities are  converted to H$\alpha$,  assuming a Balmer decrement of 2.8.  We then use 
the H$\alpha$-SFR calibration given by \cite{Kennicutt}, scaled appropriately to a \cite{Chabrier} IMF.   To derive SFRs from \oii\ luminosity, we use the calibration given by \cite{Moustakas06}, which includes an $M_B$ dependence, but does not require that the line measurement be corrected for dust.

In order to make a direct comparison to the SED-derived SFRs, we must  correct the emission line fluxes for both slit losses and and extraction aperture losses. 
 Rather than model these losses (which would not account for uncertainties in the absolute flux calibration), 
we compare our data to narrowband imaging in COSMOS.   First, we infer the emission line flux from the narrowband (NB816) photometry, using the methods outlined in 
\cite{Ly} and \cite{Takahashi}.    Since the NB816 bandpass has a non-uniform throughput, we calculate a correction to this flux based on the relative throughput at the  observed wavelengths of the emission lines.  (This correction can be important, since it is not unusual that both of the \oiii\ doublet lines are found away from the central wavelength of the NB816 filter.)  Because the COSMOS catalog contains total magnitudes (under the reasonable assumption that the galaxies are unresolved with 1.5\arcsec\ FWHM spatial resolution),  the line fluxes inferred from narrowband imaging are not subject to aperture losses.    For  17 galaxies in our sample with emission lines that are bright enough to be detected in the narrowband imaging, we can infer the aperture losses by comparing  these narrowband imaging fluxes to our spectroscopic  fluxes.      We find typical correction factors of 1.1-2.0, with a median of 1.4.     We adopt this average  correction to explore systematic differences between star formation indicators.

\begin{figure*} 
\plotone{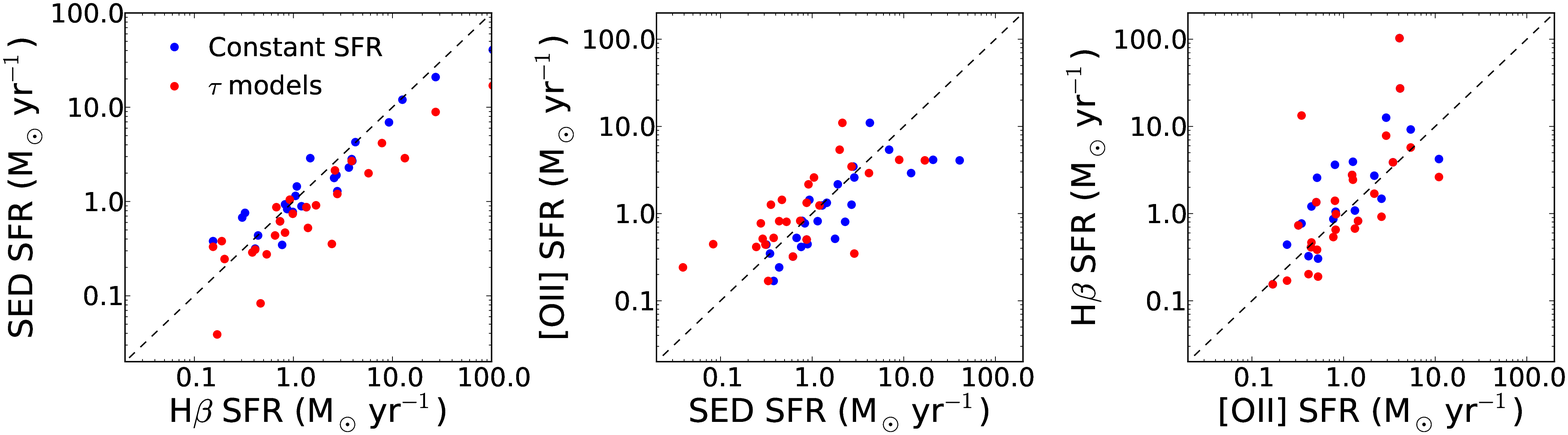} 
\caption{Star formation rates derived three different methods are compared.    Blue points assume a constant star formation history,
and the red points assume  exponentially declining SFRs ($\tau$ models) discussed in \S 3.2  The \hb-derived SFRs are corrected for dust extinction using these SED fits. (The center and right panels-- where the SED-independent \oii\ SFRs are compared-- show that the $\tau$ models tend towards slightly less dust.)    The \oii\ and \hb\ SFRs are corrected by an average factor of 1.4 to account for slit-losses.   Systematic shifts between the different indicators are at most 0.2 dex.   } 
\label{SFR_cf_fig} 
\end{figure*}

Figure \ref{SFR_cf_fig} compares the SFRs derived from the SEDs, \hb, and \oii.   In addition to the SED fits described in \S 3.2 (which use exponentially declining star formation histories with $\tau$ as a free parameter; red points), we also make the comparison for constant 
star-forming models (blue points).     This exercise shows that the SFRs derived using different methods are correlated, so that we can easily distinguish the galaxies with high and low SFRs.  However, Figure \ref{SFR_cf_fig}  also shows that there are systematic shifts as large as 0.2 dex between the different estimators.   Determining the most accurate star formation indicator is beyond the scope 
of this paper.    Ultimately, a systematic uncertainty of 0.2 dex does not affect our conclusions.  In \S \ref{fmr_sec}, we show that 
the scatter in the MZ relation is correlated with SFRs.  As with the local FMR, metallicity depends  only weakly on the SFR, so that a 0.2 dex offset in SFR translates to a 0.03 dex shift in the metallicity predicted by the local FMR \citep{Mannucci10, Mannucci11}.   This difference in metallicity is much smaller than the uncertainties on the FMR residuals shown in Figure \ref{fmrfig}.    Likewise, in \S \ref{model_sec} we use our SFR-M$_{*}$ relation to constrain models. However, 0.2 dex of systematic uncertainty is small compared to the dynamic range of SFRs  shown in Figure \ref{mz_models}.  
An systematic offset is unlikely to alter our conclusions, as the SFR-M$_{*}$ trends discussed in \S \ref{model_sec} can be seen (albeit less obviously) from the higher mass data alone. 

\begin{thebibliography} 



\bibitem[Allende Prieto et al.(2001)]{solarox} 
Allende Prieto, C., Lambert, D.~L., \& Asplund, M. 2001, \apjl, 557, L63

\bibitem[Andrews \& Martini(2013)]{AM12} 
Andrews, B.~H., \& Martini, P. 2013, \apj, 765, 140

\bibitem[Atek et al.(2011)]{Atek11} 
Atek, H., Siana, B., Scarlata, C.,  et al. 2011, \apj, 743, 121 

\bibitem[Baldwin, Phillips, \& Terlevich(1981)]{BPT} 
Baldwin, J.~A., Phillips, M.~M., \& Terlevich, R. 1981, \pasp, 93, 5

\bibitem[Berg et al.(2012)]{Berg}
Berg, D.~A., Skillman, E.~D., Marble, A.~R., et al. 2012, \apj,  754, 98 


\bibitem[Bouch\'e et al.(2010)]{Bouche}
Bouche, N.,  Dekel, A., \& Genzel, R., et al. 2010, \apj, 718, 1001


\bibitem[Brinchmann et al.(2004)]{Brinchmann04} 
Brinchmann, J., Charlot, S., White, S.~D.~M., et al. 2004, \mnras, 351, 1151

\bibitem[Brinchmann et al.(2008)]{Brinchmann08}
Brinchmann, J., Pettini, M., \& Charlot, S. 2008, \mnras, 385, 769

\bibitem[Brooks et al.(2007)]{Brooks07}
Brooks, A., Governato, F., Booth, C.~M., et al. 2007, \apj,  655, 17L



\bibitem[Bruzual \& Charlot(2003)]{BC03} 
Bruzual, G. \& Charlot, S. 2003, \mnras, 344, 1000


\bibitem[Bruzual(2007)]{Bruzual07}
Bruzual, G. 2007, ASPC, 374, 303


\bibitem[Calzetti et al.(2000)]{Calzetti}
Calzetti, D., Armus, L., Bohlin, R.~C., et al. 2000, \apj, 533, 682

\bibitem[Capak et al.(2007)]{Capak07}
Capak, P., Aussel, H., Ajiki, M.,  et al. 2007, \apjs, 172, 99

\bibitem[Cardelli et al.(1989)]{Cardelli} 
Cardelli, J.~A., Clayton, G.~C., \& Mathis, J.~S. 1989, \apj, 345, 245


\bibitem[Chabrier(2003)]{Chabrier} 
Chabrier, G. 2003, \pasp, 115, 763

\bibitem[Conroy \& Gunn(2010)]{Conroy10} 
Conroy, C., \& Gunn, J.~E. 2010, \apj, 172, 833

\bibitem[Cooper et al.(2012)]{Cooper} 
Cooper, M.~C., Newman, J.~A., Davis, M., Finkbeiner, D.~P., Gerke, B.~F. 2012, ASCL, 1203.003



\bibitem[Cowie \& Barger(2008)]{CB08}
Cowie, L.~L., \& Barger, A.~J. 2008, \apj, 686, 72

\bibitem[Cresci et al.(2010)]{Cresci10} 
Cresci, G., Mannucci, F., Maiolino, R., et al. 2010, Nature, 467, 811

\bibitem[Cresci et al.(2012)]{Cresci} 
Cresci, G., Mannucci, F., Sommariva, V., et al. 2012, \mnras, 421, 262

\bibitem[Dalcanton(2007)]{Dalcanton07}
Dalcanton, J.~J. 2007, \apj, 658, 941

\bibitem[Dav\'e et al.(2011a)]{Dave11a}
Dav\'e, R., Oppenheimer, B.~D., \&  Finlator, K.,2011, \mnras, 415, 11

\bibitem[Dav\'e et al.(2011b)]{Dave11b}
Dav\'e, R., Finlator, K., \& Oppenheimer, B.~D. 2011, \mnras, 416, 1354

\bibitem[Dav\'e et al.(2012)]{Dave12} 
Dav\'e, R., Finlator, K., \& Oppenheimer, B.~D., 2012, \mnras, 421, 98 

\bibitem[Dayal et al.(2013)]{Dayal} 
Dayal, P., Ferrara, A., \& Dunlop, J.~S. 2013,  \mnras,  430, 2891 

\bibitem[Dekel et al.(2009)]{Dekel} 
Dekel, A., Birnboin, Y., Engel, G.,  et al. 2009, Nature, 457, 451 


\bibitem[Dressler et al.(2011a)]{Dressler11} 
Dressler, A., Martin, C.~L., Henry, A., Sawicki, M., \& McCarthy, P. 2011, \apj, 740, 71 

\bibitem[Dressler et al.(2011b)]{imacs} 
Dressler, A., Bigelow, B., Hare, T.,   et al. 2011, \pasp, 123, 288 

\bibitem[Dutton et al.(2010)]{Dutton}
Dutton, A.~A., van den Bosch, F.~C., Dekel, A. 2010, \mnras, 405, 1690 

\bibitem[Edmunds(1990)]{Edmunds} 
Edmunds, M.~G. 1990, \mnras, 246, 678 


\bibitem[Elbaz et al.(2007)]{Elbaz}
Elbaz, D., Daddi, E., Le Borgne, D., et al. 2007, \aap, 468, 33


\bibitem[Ellison et al.(2008)]{Ellison08}
Ellison, S., Patton, D.~R., Simard, L., \& McConnachie, A.~W. 2008, \apj, 672, 107L

\bibitem[Erb et al.(2006)]{Erb06} 
Erb, D.~K., Shapley, A.~E., Pettini, M., et al.  2006, \apj, 644, 813

\bibitem[Erb(2008)]{Erb08} 
Erb, D.~K. 2008, \apj, 674, 151

\bibitem[Erb et al.(2012)]{Erb12} 
Erb, D.~K., Quider, A.~M., Henry, A.~L., \& Martin, C. 2012, \apj, 759, 26


\bibitem[Faber et al.(2003)]{Faber}
Faber, S.~M., Phillips, A.~C., Kibrick, R.I.,  et al. 2003, SPIE, 4841, 1657  

\bibitem[Faucher-Giguere et al.(2011)]{FG11} 
Faucher-Giguere, C.~A., Keres, D., \& Ma, C.-P. 2011, \mnras, 417, 2982

\bibitem[Fillipenko(1982)]{Fillipenko} 
Fillipenko, A.~V. 1982, \pasp, 94, 715

\bibitem[Finlator \& Dav\'e(2008)]{Finlator08} 
Finlator, K., \& Dav\'e, R. 2008, \mnras, 385, 2181

\bibitem[Finlator et al.(2011)]{Finlator11} 
Finlator, K., Oppenheimer, B.~D., Dav\'e, R. 2011, \mnras, 410, 1730


\bibitem[Hainline et al.(2009)]{Hainline09} 
Hainline, K.~N., Shapley, A.~E., Kornei, K.~A., et al. 2009, \apj, 701, 52

\bibitem[Heckman et al.(1990)]{Heckman90} 
Heckman, T.~M., Armus, L., \& Miley, G.~K. 1990, \apjs, 74, 833


\bibitem[Henry et al.(2012)]{Henry12} 
Henry, A.~L., Martin, C.~L., Dressler, A., Sawicki, M., \& McCarthy, P. 2012, \apj, 744, 149 

\bibitem[Henry et al.(2007)]{Henry07} 
Henry, A.~L., Turner, J.~L., Beck, S.~C., Crosthwaite, L.~P., \& Meier, D.~S. 2007, \aj, 133, 757

\bibitem[Hopkins et al.(2012)]{Hopkins} 
Hopkins, P.~F., Quataert, E., \& Murray, N. 2012, \mnras, 421, 3522

\bibitem[Hu et al.(2009)]{Hu09}
Hu, E., Cowie, L.~L., Kakazu, Y. \& Barger, A. 2009, \apj, 698, 2014

\bibitem[Ilbert et al.(2009)]{Ilbert09} 
Ilbert, O.,  Capak, P., Salvato, M., et al. 2009, \apj, 690, 1236

\bibitem[Juneau et al.(2011)]{Juneau11} 
Juneau, S., Dickinson, M., Alexander, D.~M., \& Salim, S. 2011, \apj, 736, 104

\bibitem[Kakazu et al.(2007)]{Kakazu} 
Kakazu, Y., Cowie, L.~L., \& Hu, E.~M. 2007, \apj, 668, 853

\bibitem[Kashikawa et al.(2011)]{Kashikawa11} 
Kashikawa, N.,  Shimasaku, K., Matsuda, Y., et al. 2011, \apj, 648, 7 

\bibitem[Kewley \& Dopita(2002)]{KD02} 
Kewley, L.~J., \& Dopita, M.~A. 2002, \apjs, 142, 35

\bibitem[Kewley \& Ellison(2008)]{KE08} 
Kewley, L.~J., \& Ellison, S.~L. 2008, \apj, 681, 1183

\bibitem[Kennicutt(1998)]{Kennicutt} 
Kennicutt, R.~C., 1998, \araa, 36, 189

\bibitem[Kirby et al.(2011)]{Kirby11} 
Kirby, E.~N., Martin, C.~L., \& Finlator, K. 2011, \apj, 742, 25L


\bibitem[Kniazev et al.(2003)]{Kniazev} 
Kniazev, A.~Y., Grebel, E.~K., Hao, L., et al. 2003, \apj, 593, 73L

\bibitem[Kobulnicky \& Kewley(2004)]{KK04} 
Kobulnicky, H.~A., \& Kewley, L.~J. 2004, \apj, 617, 240

\bibitem[Kornei et al.(2012)]{Kornei12} 
Kornei, K.~A., Shapley, A.~E., Martin, C.~L., et al. 2012, \apj, 758, 135 

\bibitem[Kriek et al.(2009)]{FAST}
Kriek, M., van Dokkum, P.~G., Labb\'e, I., et al.  2009, \apj, 700, 221

\bibitem[Kriek et al.(2010)]{Kriek10}
Kriek, M.,  Labb\'e, I., Conroy, C., et al. 2010, \apj, 722,  L64

\bibitem[Kroupa(2001)]{Kroupa} 
Kroupa, P. 2001, \mnras, 322, 231

\bibitem[Lamareille et al.(2009)]{Lamareille} 
Lamareille, F.,  Brinchmann, J., Contini, T., et al. 2009, \aap, 495, 53 

\bibitem[Lara-Lopez et al.(2010)]{LL10} 
Lara-L\'opez, M. A., Cepa, J., Bongiovanni, A., et al. 2010, \aap, 521, 53L 

\bibitem[Lee et al.(2006)]{Lee} 
Lee, H., Skillman, E.~D., Cannon, J.~M., et al. 2006, \apj, 647, 970

\bibitem[Le F\`evre et al.(2004)]{vvds} 
Le F\`evre, O., Vettolani, G., Paltani, S., et al. 2004, \aap, 428, 1043

\bibitem[Liang et al.(2007)]{Liang} 
Liang, Y.~C., Hammer, F., \& Yin, S.~Y.\ 2007, \aap, 474, 807 


\bibitem[Lilly et al.(2003)]{Lilly03} 
Lilly, S.~J., Carollo, C.~M., \& Stockton, A.~N. 2003, \apj, 597, 730

\bibitem[Lilly et al.(2007)]{Lilly} 
Lilly, S.~J.,  Le F\`evre, O., Renzini, A.,  et al. 2007, \apjs, 172, 70

\bibitem[Lilly et al.(2013)]{Lilly13} 
Lilly, S.~J., Carollo, M.~C., Pipino, A., Renzini, A., \& Peng, Y. 2013, arXiv:1303.5059 

\bibitem[L\'opez-S\'anchez et al.(2012)]{LS12} 
L\'opez-S\'anchez, \'A.~R. , Dopita, M.~A., Kewley, L.~J., et al. 2012, \mnras,  426, 2630 


\bibitem[Ly et al.(2007)]{Ly} 
Ly, C., Malkan, M.~A., Kashikawa, N., et al. 2007, \apj, 657, 738


\bibitem[Maiolino et al.(2008)]{Maiolino08} 
Maiolino, R., et al. 2008, \aap, 488, 463

\bibitem[Mannucci et al.(2010)]{Mannucci10} 
Mannucci, F., Cresci, G., Maiolino, R., Marconi, A., \& Gnerucci, A. 2010, \mnras, 408, 2115

\bibitem[Mannucci et al.(2011)]{Mannucci11} 
Mannucci, F., Salvaterra, R., \& Campisi, M.~A. 2011, \mnras, 414, 1263

\bibitem[Maraston(2005)]{M05} 
Maraston, C. 2005, \mnras, 362, 799

\bibitem[Martin et al.(2008)]{Martin08} 
Martin, C.~L., Sawicki, M., Dressler, A., \& McCarthy, P. 2008, \apj, 679, 942

\bibitem[Martin(2005)]{Martin05} 
Martin, C.~L. 2005, \apj, 621, 227 

\bibitem[Massey \& Gronwall(1990)]{MG90}
Massey, P., \& Gronwall, C. 1990, \apj, 358, 344 

\bibitem[McGaugh(1991)]{McGaugh} 
McGaugh, S. 1991, \apj, 380, 140

\bibitem[Moster et al.(2010)]{Moster}
Moster, B.~P., Somerville, R.~S., Maulbetsch, C., et al. L. 2010, \apj, 710, 903 

\bibitem[Moustakas et al.(2006)]{Moustakas06}
Moustakas, J., Kennicutt, R.~C., \& Tremonti, C.~A. 2006, \apj, 642, 775

\bibitem[Moustakas et al.(2011)]{Moustakas11} 
Moustakas, J., Zaritsky, D., Brown, M.,  et al. 2011, arXiv:1112.3300 

\bibitem[Murray et al.(2005)]{Murray} 
Murray, N., Quatert, E., \& Thompson, T.~A. 2005, ApJ, 618, 569

\bibitem[Nagao et al.(2006)]{Nagao} 
Nagao, T., Maiolino, R., \& Marconi, A., 2006, \aap, 459, 85


\bibitem[Newman et al.(2012)]{Newman} 
Newman, J.~A., Cooper, M.~C., Davis, M.,  et al. 2012, arXiv:1203.3192


\bibitem[Nicholls et al.(2012)]{Nicholls}
Nicholls, D., Dopita, M.~A., \& Sutherland, R.~S., \apj, 2012, 752, 148 

\bibitem[Noeske et al.(2007)]{Noeske} 
Noeske, K.~G.,  Faber, S.~M., Weiner, B.~J., et al. 2007, \apj, 660, L47


\bibitem[Okamoto et al.(2008)]{Okamoto} 
Okamoto, T., Gao, L., Theuns, T. 2008, \mnras, 390, 920

\bibitem[Oke(1990)]{Oke} 
Oke, J.~B. 1990, \aj, 99, 1621


\bibitem[Oppenheimer et al.(2009)]{Opp09} 
Oppenheimer, B.~D., Dav\'e, R., \& Finlator, K. 2009, \mnras, 396, 729 


\bibitem[Pagel et al.(1979)]{Pagel} 
Pagel, B.~E.~J., Edmunds, M.~G., Blackwell, D.~E., Chun, M.~S., \& Smith, G. 1979, \mnras, 189, 95

\bibitem[Peeples et al.(2009)]{Peeples09} 
Peeples, M.~S., Pogge, R.~W., \& Stanek, K.~Z. 2009, \apj, 695, 259

\bibitem[Peeples \& Shankar(2011)]{PS11} 
Peeples, M.~S., \& Shankar, F. 2011, \mnras, 417, 2962

\bibitem[Peimbert \& Costero(1969)]{PC69}
Peimbert, M., \& Costero, R. 1969, Bol. Obs. Ton. y Tac., 5, 3

\bibitem[Pettini \& Pagel(2004)]{PP04} 
Pettini, M., \& Pagel, B.~E.~J. 2004, \mnras, 384, 59L

\bibitem[Pilyugin(2001)]{P01}
Pilyugin, L.~S. 2001, \aap, 374, 412

\bibitem[Pilyugin \& Thuan(2005)]{P05}
Pilyugin, L.~S., \& Thuan, T.~X. 2005, \apj, 631, 231

\bibitem[Pirzkal et al.(2012)]{Pirzkal12}
Pirzkal, N.,  Rothberg, B., Ly, C., et al. 2012, \apj,  arXiv:1208.5535

\bibitem[Savaglio et al.(2005)]{Savaglio} 
Savaglio, S., Glazebrook, K., Le Borgne, D., et al. 2005, \apj, 635, 260 

\bibitem[Schaerer \& de Barros(2009)]{SdB} 
Schaerer, D.,  \& de Barros, S. 2009, \aap, 502, 423 


\bibitem[Shapley et al.(2005)]{Shapley05}
Shapley, A.~E., Coil, A.~L., Ma, C.-P.,  Bundy, K. 2005, \apj, 635, 1006

\bibitem[Soto et al.(2012)]{Soto12} 
Soto, K.~T., Martin, C.~L., Prescott, M.~K.~M., \& Armus, L. 2012, \apj, 757, 86

\bibitem[Stasi\'nska(2005)]{Stasinska} 
Stasi\'nska, G. 2005, \aap, 434, 507

\bibitem[Takahashi et al.(2007)]{Takahashi} 
Takahashi,  M.~I.,  Shioya, Y., Taniguchi, Y., et al. 2007, \apjs, 172, 456



\bibitem[Tremonti et al.(2004)]{Tremonti}
Tremonti, C.~A., Heckman, T.~M., Kauffman, G., et al. 2004, \apj, 613, 898

\bibitem[van den Bergh(1962)]{vdB62} 
van den Bergh, S. 1962, \aj, 67, 486


\bibitem[van Zee et al.(2006)]{vanZee} 
van Zee, L., Skillman, E.~D., \& Haynes, M.~P. 2006, \apj, 637, 269 

\bibitem[Weinmann et al.(2012)]{Weinmann} 
Weinmann, S.~M., Pasquali, A., Oppenheimer, B.~D., et al. 2012, \mnras, 426, 2797

\bibitem[Wofford et al.(2013)]{Wofford} 
Wofford, A., Leitherer, C., \& Salzer, J. 2013, ApJ, 765, 118 

\bibitem[Wuyts et al.(2012)]{Wuyts} 
Wuyts, E., Rigby, J.~R., Sharon, K., \& Gladders, M.~D. 2012, \apj, 755, 73

\bibitem[Xia et al.(2012)]{Xia} 
Xia, L., Malhotra, S., Rhoads, J., et al. 2012, \aj, 144, 28

\bibitem[Yates et al.(2012)]{Yates} 
Yates, R.~M., Kauffman, G., \& Guo, Q. 2012, \mnras,  422, 215 

\bibitem[Zahid et al.(2011)]{Zahid11} 
Zahid, H.~J., Kewley, L.~J.,  \& Bresolin, F. 2011, \apj, 730, 137 

\bibitem[Zhao et al.(2010)]{Zhao} 
Zhao, Y., Gao, Y., \& Gu, Q. 2010, \apj, 710, 663

\bibitem[Zibetti et al.(2013)]{Zibetti}
Zibetti, S. Gallazzi, A., Charlot, S. Pierini, D.,  \& Pasquali, A. 2013, \mnras,  428, 1479

\end{thebibliography}
\end{document}